\documentclass[preprint,3p,a4paper]{elsarticle}

\usepackage{mathrsfs}
\usepackage{amsmath}
\usepackage{stmaryrd}
\usepackage{bbding}
\usepackage{dcolumn}
\usepackage{graphicx}
\usepackage{amsfonts}
\usepackage{amssymb}
\usepackage{psfrag}
\usepackage{wrapfig}
\usepackage{subfigure}
\usepackage{makeidx}
\usepackage{bm}
\usepackage{epsf}
\usepackage{epsfig}
\usepackage{setspace}
\usepackage{graphicx}
\usepackage{epstopdf}
\usepackage{psfrag}
\usepackage{subfigure}
\usepackage{booktabs}
\usepackage{color}
\usepackage{comment}
\usepackage{ulem}
\usepackage{algorithm}
\usepackage{algorithmic}

\newtheorem{rmk}{Remark}

\renewcommand{\vec}[1]{{\boldsymbol{#1}}}

\begin{document}
	
	\title{Unified gas-kinetic wave-particle method for frequency-dependent radiation transport equation}
	
	\author[HKUST1]{Xiaojian Yang}
	\ead{xyangbm@connect.ust.hk}
	
	\author[HKUST2]{Yajun Zhu}
	\ead{mazhuyajun@ust.hk}
	
	\author[IAPCM]{Chang Liu}
	\ead{liuchang@iapcm.ac.cn}
	
	\author[HKUST1,HKUST2,HKUST3]{Kun Xu\corref{cor1}}
	\ead{makxu@ust.hk}
	
	\address[HKUST1]{Department of Mechanical and Aerospace Engineering, The Hong Kong University of Science and Technology, Clear Water Bay, Kowloon, Hong Kong, China}
	\address[HKUST2]{Department of Mathematics, The Hong Kong University of Science and Technology, Clear Water Bay, Kowloon, Hong Kong, China}
	\address[HKUST3]{Shenzhen Research Institute, The Hong Kong University of Science and Technology, Shenzhen, China}
	\address[IAPCM]{Institute of Applied Physics and Computational Mathematics, Beijing, China}
	
	\cortext[cor1]{Corresponding author}
	
	\begin{abstract}
		The multi-frequency radiation transport equation (RTE) system models the photon transport and the energy exchange process between the background material and different frequency photons.
		In this paper, the unified gas-kinetic wave-particle (UGKWP) method for multi-frequency RTE is developed to
		capture the multiscale non-equilibrium transport in different optical regimes.
		In the UGKWP, a multiscale evolution process is properly designed to obtain both non-equilibrium transport in the optically thin regime and thermal diffusion process in the optically thick regime automatically.
		At the same time, the coupled macroscopic energy equations for the photon and material are solved implicitly by the matrix-free source iteration method.
		With the wave-particle decomposition strategy,
		the UGKWP method has a dynamic adaptivity for different regime physics.
		In the optically thick regime,
		no particles will be sampled in the computational domain due to the intensive energy exchange between photon and background material,
		and the thermal diffusion solution for the photon transport will be recovered.
		While in the optically thin regime,
		stochastic particles will play a dominant role in the evolution
		and the non-equilibrium free transport of photon is automatically followed.
		In the frequency-dependent transport,
		the frequency carried by the simulating particle will be determined by a linear-frequency sampling strategy.
		In addition, to better resolve the sharp transition of opacity in the photon transport across a cell interface,
		the free streaming time of simulating particle in the UGKWP method will be reset when it passes through the interface.
		Moreover, the numerical relaxation time is properly defined to increase the particle proportion in the sharp opacity transition region 
		in order to avoid numerical oscillation.
		Several typical test cases for the RTE system have been calculated to demonstrate the accuracy and reliability of the current frequency-dependent UGKWP method.
	\end{abstract}
	
	\maketitle
	
	\section{Introduction}
	Radiation plays important roles in high-energy-density physics, such as star formation, inertial confinement fusion (ICF), etc \cite{rad-book-MC-vassiliev2017monte}.
	The radiation transport equation (RTE) models the events of photons in the material medium,
	such as the transport, scattering, absorption, and black-body emission, etc.
	RTE is capable of describing the multi-scale features of photon's behaviors,
	which is mainly determined by the local opacity of the material medium,
	reflecting the probability of photons being absorbed.
	For example, if the opacity is much smaller, the photon will transport for a long distance and does not interact with the background material.
	In other words, the material is very transparent for the radiation.
	In contrast, in the zones with a large opacity, photons will interact with the material frequently, such as the collision and absorption.
	Also, some photons will be emitted by the material medium,
	and the distribution of emitted photons follows the local thermodynamic equilibrium, i.e., Planck's function.
	In addition, the absorption and emission processes of photons are strongly coupled with the exchange between the radiation energy and material internal energy.
	Due to the multi-scale property of photons' transport behaviors and the complexity of interaction with background media,
	it is nearly impossible to obtain the analytical solution for practical problems.
	Therefore, numerical simulation plays an indispensable role, and developing advanced numerical methods for RTE is necessary.
	
	In the past decades, lots of attempts have been made to develop a numerical method with high accuracy and efficiency.
	Generally, the methods can be classified into two groups, the deterministic method and the stochastic method.
	The spherical harmonic $\left(P_N\right)$ method is one of the representative deterministic methods.
	In the $P_N$ method, the distribution of radiation intensity $\left(I\right)$ is described based on the spherical harmonic basis functions,
	and further a system of equations for $N$-th order expression of $I$ will be derived and solved \cite{rad-PN-filter-mcclarren2010robust, rad-PN-filter-mcclarren2010simulating}.
	Another widely-used deterministic method is the discrete ordinates $\left(S_N\right)$ method.
	In the $S_N$ method, the continuous angular space is represented by finite discrete angular points.
	One shortcoming of the $S_N$ method is the so-called "ray effect",
	and the underlying causes and associated remedies have been studied, \cite{rad-SN-ray-effect-lathrop1971remedies, rad-SN-ray-effect-coelho2001role, rad-SN-ray-effect-cumber2000ray, rad-SN-ray-effect-zhu2020ray}.
	
	For the stochastic method, the implicit Monte Carlo (IMC) method proposed by Fleck and Cummings in the 1970s is the most popular one, and widely used in the community of high-energy-density physics \cite{rad-IMC-first-fleck1971implicit}.
	Different from the (explicit) Monte Carlo method, effective scattering is adopted to replace part of absorption-emission processes,
	which occur frequently, especially in the optically thick regime.
	As a result, a large time step can be employed to improve the computation efficiency.
	Accordingly, a commonly known drawback of IMC is the so-called teleportation error,
	which indicates the wave or photons transport faster.
	Lots of studies have been done to further improve the accuracy of IMC.
	For example, source titling is one popular technique to improve the sub-cell resolution and thus reduce the teleportation error \cite{rad-IMC-source-title-wollaeger2016implicit}.
	Recently, the strategy adopting two kinds of particles, one for photons and another for material,
	has been employed in the semi-analog Monte Carlo (SMC) method,
	as well as its further version, implicit semi-analog Monte Carlo (ISMC) method \cite{rad-IMC-SMC-ahrens2001semi, rad-IMC-ISMC-poette2020new, rad-mf-IMC-ISMC-steinberg2022multi}.
	In SMC and ISMC, the material particle will be transformed from/to the radiation particle (photon) due to absorption/emission,
	by which the teleportation problem is attenuated.
	A similar idea is also taken in the discrete implicit Monte Carlo (DIMC) method,
	which can further reduce statistical noises \cite{rad-IMC-DIMC-steinberg2022new}.
	In the limiting optically thick regime,
	both the deterministic methods and the stochastic particle methods should recover the solution of the equilibrium diffusion equation,
	which can be obtained by asymptotic analysis \cite{rad-AP-diffusion-larsen1983asymptotic}.
	Accordingly, the hybrid method couples the radiation transport equation and diffusion equation so as to improve the computational efficiency in the zone with high opacity.
	Generally, the standard IMC is employed in optically thin regions,
	while the diffusion approximation is taken in optically thick regions to replace the tremendous amount of MC steps,
	such as discrete diffusion Monte Carlo (DDMC) \cite{rad-hybrid-DDMC-densmore2007hybrid, rad-mf-hybrid-DDMC-densmore2012hybrid},
	implicit Monte Carlo diffusion (IMD) \cite{rad-hybrid-IMCD-gentile2001implicit, rad-mf-hybrid-IMCD-cleveland2010extension},
	implicit Monte Carlo coupling random walk (RM) method \cite{rad-hybrid-random-walk-giorla1987random}, etc.
	As with most hybrid methods, getting a proper domain decomposition for different regimes is a tough issue \cite{rad-hybrid-interface-densmore2006interface}.
	
	The idea of direct modeling in the unified gas-kinetic scheme (UGKS) makes it capable of solving the multi-scale transport problems \cite{UGKS-xu2010unified, UGKS-book-framework-xu2021cambridge}. In recent years, the asymptotic preserving unified gas-kinetic scheme (APUGKS) for grey radiation has been proposed, and further developed systematically \cite{rad-UGKS-sun2015asymptotic, rad-mf-UGKS-sun2015asymptotic, rad-UGKS-unstruct-sun2017multidimensional, rad-UGKS-hydro-sun2020multiscale}.
	Specifically, the macroscopic system is solved by source iteration,
	and the UGKS provides the multi-scale numerical flux based on the integral solution of RTE.
	In the optically thick limiting, the solution of the diffusion equation in \cite{rad-AP-diffusion-larsen1983asymptotic} will be recovered automatically.
	To describe non-equilibrium distribution more efficiently,
	a particle version of UGKS, i.e., the unified gas-kinetic particle (UGKP) method is developed for the grey radiation \cite{rad-kp-shi2020asymptotic, rad-kp-shi2021improved, rad-mf-kp-li2023unified, rad-mf-kp-hu2023unified}.
	Moreover, the unified gas-kinetic wave-particle (UGKWP) method based on the wave-particle adaptation has been developed to further improve the efficiency of the UGKP method,
	and it has been used in rarefied flow, plasma, and multi-phase flow, etc \cite{WP-first-liu2020unified, WP-second-zhu-unstructured-mesh-zhu2019unified, WP-four-liu2020unified, WP-six-gas-particle-yang2021unified, WP-gasparticle-fluidized-yang2023unified}.
	Compared to UGKS, the UGKWP method has greater advantages for radiation transport problems,
	such as the self-adaption of discrete velocity space (mainly for the optically thin regime),
	the lower requirement of memory (mainly for the optically thick regime), etc.
	As pioneering works, the UGKWP method based on the operator splitting strategy and an implicit UGKWP (IUGKWP) method based on the source iteration strategy,
	have been proposed for grey radiation \cite{rad-wp-li2020unified, rad-wp-implicit-liu2023implicit}.
	The results present a significant efficiency improvement compared to the IMC method, especially for the IUGKWP.
	
	In the grey radiation model, the opacity is independent of the photon's frequency.
	As a result, the flow regime is mainly determined by the local opacity of the material medium.
	In reality, the photon's absorption and emission coefficients highly depend on its frequency,
	which requires taking consideration of the photon's frequency in the numerical scheme for more accurate prediction.
	For example, the material is more transparent for the high-frequency photons than the low-frequency ones \cite{rad-mf-UGKS-sun2015asymptotic, rad-mf-kp-li2023unified}.
	Generally, the photon's multi-frequency feature increases the complexity of solving the RTE,
	which brings more challenges for numerical methods, especially for the hybrid methods \cite{rad-mf-hybrid-DDMC-densmore2012hybrid, rad-mf-hybrid-IMCD-cleveland2010extension}.
	However, due to the inherent adaptation of wave and particle decomposition based on the flow regime, the UGKWP method is suitable for the multi-frequency case. Specifically, based on the multi-group strategy for the frequency-dependent case, each group can find the most optimal wave-particle distribution.
	In this paper, for a more realistic model, the UGKWP method for the frequency-dependent RTE system will be developed.
	
	This paper is organized as follows.
	Section 2 introduces the governing equations of the frequency-dependent RTE system,
	and gives the details about the multi-scale UGKWP method for solving the frequency-dependent RTE system.
	Section 3 shows the numerical results.
	The last section is the conclusion.
	
	\section{Methodology}
	In this section, the governing equations of the frequency-dependent RTE system will be introduced first,
	and then the multi-scale UGKWP method under the finite volume framework will be presented in detail.
	
	\subsection{Frequency-dependent RTE system}
	For the frequency-dependent radiation transfer system, the governing equation for radiation intensity can be written as
	\begin{equation} \label{eq:mfGovI}
	\frac{\epsilon^2}{c}\frac{\partial I}{\partial t}
	+ \epsilon \vec{\Omega} \cdot \nabla_{\vec{x}} I
	= \sigma \left(B\left(\nu,T\right)-I\right),
	\end{equation}
	together with the energy balance equation between radiation and material energy
	\begin{equation}
	\epsilon^2 C_v \frac{\partial T}{\partial t} \equiv \epsilon^2 \frac{\partial U}{\partial t} =
	\int_{0}^{\infty} \int_{\mathcal{S}^2}
	\sigma\left[I\left(\vec{\Omega}, \nu\right) -  B\left(\nu, T\right)\right]
	\text{d}\vec{\Omega} \text{d}\nu
	\end{equation}
	where $I\left(\vec{x},t,\vec{\Omega},\nu\right)$ is the specific radiation intensity,
	$\vec{\Omega}$ is space angle,
	$\nu$ is the photon frequency,
	$\sigma$ denotes the opacity,
	$c$ is the speed of light,
	$\epsilon$ is the Knudsen number,
	$C_v$ is the heat capacity,
	and $U$ is the material energy density.
	Note that the scattering and internal source terms are omitted in Eq.~\eqref{eq:mfGovI}.
	The emitted photon from the background material follows the Planck distribution, i.e.,
	\begin{equation*}
	B\left(\nu, T\right) = \frac{2h\nu^3}{c^2} \frac{1}{e^{h\nu/kT}-1},
	\end{equation*}
	which satisfies
	\begin{equation*}
	\int_{0}^{\infty} \int_{\mathcal{S}^2} B\left(\nu, T\right) \text{d}\vec{\Omega} \text{d}\nu = ac T^4,
	\end{equation*}
	where $h$ is the Planck constant,
	$k$ is the Boltzmann constant,
	and $a$ is the radiation constant
	\begin{equation*}
	a = \frac{8 \pi^5 k^4}{15 h^3 c^3}.
	\end{equation*}
	
	In this study, the multigroup method is used for the frequency-dependent RTE.
	Specifically, the frequency space $\left(0,\infty\right)$ is divided into $G$ intervals,
	and the $g$-th group covers the frequencies $\nu\in[\nu_{g-1/2}, \nu_{g+1/2}]$ with $g=1,...,G$.
	Integrate Eq.~\eqref{eq:mfGovI} over the $g$-th group,
	\begin{equation}
	\int_{\nu_{g-1/2}}^{\nu_{g+1/2}} \left[ \frac{\epsilon^2}{c}\frac{\partial I}{\partial t}
	+ \epsilon \vec{\Omega} \cdot \nabla_\vec{x} I \right] \text{d}\nu
	= \int_{\nu_{g-1/2}}^{\nu_{g+1/2}} \left[ \sigma \left(B\left(\nu,T\right)-I\right) \right] \text{d}\nu,
	\end{equation}
	which leads to
	\begin{equation}
	\frac{\epsilon^2}{c}\frac{\partial I_g}{\partial t}
	+ \epsilon \vec{\Omega} \cdot \nabla_\vec{x} I_g
	= \sigma_g^e B_g - \sigma_g^a I_g,
	\end{equation}
	where the specific radiation intensity of $g$-th group denoted by $I_g$ is defined as
	\begin{equation*}
	I_g\left(\vec{x},t,\vec{\Omega}\right) = \int_{\nu_{g-1/2}}^{\nu_{g+1/2}} I\left(\vec{x},t,\vec{\Omega},\nu\right) \text{d}\nu.
	\end{equation*}
	The equivalent emission of $g$-th group denoted by $B_g$ is defined as
	\begin{equation*}
	B_g\left(T\right) = \int_{\nu_{g-1/2}}^{\nu_{g+1/2}} B\left(\nu,T\right) \text{d}\nu.
	\end{equation*}
	$\sigma_g^e$ and $\sigma_g^a$ satisfy
	\begin{equation*}
	\sigma_g^e B_g
	= \int_{\nu_{g-1/2}}^{\nu_{g+1/2}} \sigma B\left(\nu,T\right) \text{d}\nu,
	\end{equation*}
	and
	\begin{equation*}
	\sigma_g^a I_g
	= \int_{\nu_{g-1/2}}^{\nu_{g+1/2}} \sigma I \text{d}\nu,
	\end{equation*}
	respectively.
	It should be noted that $\sigma_g^a$ can not be obtained directly due to the unknown $I$.
	In the current study, we take $\sigma_g^a = \sigma_{g}^e = \sigma_{g}\left(\nu\right)$,
	and the frequency-dependent opacity of $g$-th group $\sigma_{g}\left(\nu\right)$ will be given for specific problems.
	
	Now, the multi-group RTE system can be written as
	\begin{equation*}
	\begin{cases}
	\frac{\epsilon^2}{c}\frac{\partial I_g}{\partial t}
	+ \epsilon \vec{\Omega} \cdot \nabla_\vec{x} I_g
	= \sigma_g B_g - \sigma_g I_g, \quad g=1,...,G, \\
	\epsilon^2 C_v \frac{\partial T}{\partial t}
	= \sum\limits_{g=1}^{G} \int_{\mathcal{S}^2} \left( \sigma_g I_g - \sigma_g B_g \right)\text{d}\vec{\Omega}.
	\end{cases}
	\end{equation*}
	Integrating the above RTE system in angular space $\mathcal{S}^2$,
	the associated governing equations of the macroscopic variables can be obtained
	\begin{equation*}
	\begin{cases}
	\frac{\partial \rho_g}{\partial t}
	+ \nabla_\vec{x} \cdot \vec{F}_g
	= \frac{c \sigma_g}{\epsilon^2} 4\pi B_g
	- \frac{c \sigma_g}{\epsilon^2} \rho_g, \quad g=1,...,G, \\
	C_v \frac{\partial T}{\partial t} = \sum\limits_{g=1}^{G} \frac{1}{\epsilon^2}\left(\sigma_g \rho_g - \sigma_g 4\pi B_g \right),
	\end{cases}
	\end{equation*}
	where $\rho_g = \int_{\mathcal{S}^2} I_g\left(\vec{\Omega}\right) \text{d}\vec{\Omega}$ is the radiative energy density for the $g$-th group,
	and $\vec{F}_g = \langle \frac{c}{\epsilon} \vec{\Omega} I_g \rangle = \int_{\mathcal{S}^2} \frac{c}{\epsilon} \vec{\Omega} I_g \left(\vec{\Omega}\right) \text{d}\vec{\Omega}$ is the radiation energy flux.
	
	\subsection{Multi-scale UGKWP method}
	The UGKWP method is a finite volume method based on the conservation laws.
	For a discrete cell $i$,
	the macroscopic RTE system about the flow variables $\rho_{g}$ and $T$ is discretized in an implicit way by
	\begin{equation}\label{macroEquImplicit}
	\begin{cases}
	\rho_{g,i}^{n+1} = \rho_{g,i}^n - \frac{\Delta t}{V_i} \sum\limits_{S_{ij}\in \partial V_i}F_{g,ij}S_{ij} + \frac{c \Delta t}{\epsilon^2} \left(\sigma_{g,i}^{n+1} 4\pi B_{g,i}^{n+1} - \sigma_{g,i}^{n+1} \rho_{g,i}^{n+1} \right), \\
	C_vT_i^{n+1} = C_vT_i^{n} + \sum\limits_{g=1}^{G} \frac{\Delta t}{\epsilon^2}\left(\sigma_{g,i}^{n+1} \rho_{g,i}^{n+1} - \sigma_{g,i}^{n+1} 4\pi B_{g,i}^{n+1} \right),
	\end{cases}
	\end{equation}
	where $V_i$ is the volume of cell $i$,
	$\partial V_i$ denotes the set of interfaces of cell $i$,
	$S_{ij}$ is the area of the $j$-th interface of cell $i$.
	$\rho_{g,i}$ and $T_i$ are the associated cell-averaged variables.
	$F_{g,ij}$ denotes the macroscopic flux across the interface $S_{ij}$,
	which can be written as
	\begin{equation*}
	F_{g,ij} =\frac{1}{\Delta t}\int_{0}^{\Delta t} \int_{\mathcal{S}^2} \frac{c}{\epsilon}\vec{\Omega}\cdot\vec{n}_{ij} I_{g,ij} \text{d}\vec{\Omega}\text{d}t,
	\end{equation*}
	where $\vec{n}_{ij}$ denotes the normal vector of interface $S_{ij}$ pointing to the outside of cell $i$,
	$I_{g,ij}$ is the time-dependent specific radiation intensity on the interface $S_{ij}$.
	
	The above implicit formula
	will be solved based on source iteration,
	which will be introduced in detail later.
	UGKWP method will be employed to provide the multi-scale flux. So, the following mainly introduces how to obtain the time-dependent $I_g$ at the cell interface and further calculate the flux $\vec{F}_g$ in the UGKWP method.

	For the above macroscopic system \eqref{macroEquImplicit},
	the essential point to construct a multi-scale flux $F_{g,ij}$,
	which could recover the non-equilibrium physics of photons in all flow regimes.
	As same as in the UGKS, this is carried out by the integral solution of the RTE
	\begin{equation*}
	\frac{\partial I_g}{\partial t}
	+ \frac{c}{\epsilon} \vec{\Omega} \cdot \nabla_\vec{x} I_g
	= \frac{c \sigma_g}{\epsilon^2} \left(B_g-I_g\right),
	\end{equation*}
	along the characteristic line $\vec{x}'=\vec{x}+\frac{c}{\epsilon}\vec{\Omega}(t'-t)$, i.e.,
	\begin{equation}\label{integSolI}
	I_g(\vec{x},t,\vec{\Omega})=\frac{1}{\epsilon^2/c\sigma_g}\int_{t_0}^t B_g(\vec{x}',t',\vec{\Omega})e^{-\frac{t-t'}{\epsilon^2/c\sigma_g}}\text{d}t'
	+e^{-\frac{t-t_0}{\epsilon^2/c\sigma_g}}I_{g,0}(\vec{x}-\frac{c}{\epsilon}\vec{\Omega}\left(t-t_0\right), \vec{\Omega}),
	\end{equation}
	where $I_{g,0}$ is the initial distribution at time $t_0$,
	and $B_g$ denotes the local equilibrium state.
	For brevity, we assume $t_0=0$ and $\vec{x}=\vec{0}$ to indicate the start of each time step and the center of the investigated cell interface.
	Then Eq.~\eqref{integSolI} can be written as
	\begin{equation*}
	\begin{aligned}
	I_g(\vec{0},t,\vec{\Omega}) & =
	\frac{1}{\epsilon^2/c\sigma_g} \int_{0}^t B_g(\vec{x}',t',\vec{\Omega}) e^{-\frac{t-t'}{\epsilon^2/c\sigma_g}} \text{d}t'
	+e^{-\frac{t}{\epsilon^2/c\sigma_g}} I_{g,0}(-\frac{c}{\epsilon}\vec{\Omega}t, \vec{\Omega}), \\
	& \overset{def}{=} I^{eq}_g(\vec{0},t,\vec{\Omega}) + I^{fr}_g(\vec{0},t,\vec{\Omega}).
	\end{aligned}
	\end{equation*}
	The time-averaged macroscopic flux across the interface $S_{ij}$ during one time step $\left[0,\Delta t\right]$ for the $g$-th group can be constructed by taking moments of the time-dependent $I_g\left(t\right)$.
	Specifically, we have
	\begin{equation*}
	F_{g,ij}= \frac{1}{\Delta t} \int_{0}^{\Delta t} \int_{\mathcal{S}^2} \frac{c}{\epsilon}\vec{\Omega}\cdot\vec{n}_{ij}
	\left[I^{eq}_g(\vec{0},t,\vec{\Omega}) + I^{fr}_g(\vec{0},t,\vec{\Omega})\right]
	\text{d}\vec{\Omega}\text{d}t \overset{def}{=}
	F^{eq}_{g,ij}+F^{fr}_{g,ij}.
	\end{equation*}
	
	Considering second-order accuracy,
	$B_g$ is assumed to have a linear distribution in space and time,
	then we can obtain the equilibrium part of $I_g(t)$ by
	\begin{align}\label{IeqFormula}
	I^{eq}(\textbf{0},t,\vec{\Omega})
	= c_1 \frac{1}{4\pi} B_g\left(\textbf{0},\vec{\Omega}\right)
	+ c_2 \frac{1}{4\pi} \frac{c}{\epsilon}\vec{\Omega} \cdot B_{g,\vec{x}}\left(\textbf{0},\vec{\Omega}\right)
	+ c_3 \frac{1}{4\pi} B_{g,t}\left(\textbf{0},\vec{\Omega}\right),
	\end{align}
	where $B_{g,\vec{x}}$ and $B_{g,t}$ are the spatial and temporal derivatives of $B_g$, respectively.
	The related coefficients are
	\begin{align*}
	c_1 & = 1 - e^{-\frac{t}{\epsilon^2/c\sigma}},                                                                   \\
	c_2 & = \left(t+\frac{\epsilon^2}{c\sigma}\right) e^{-\frac{t}{\epsilon^2/c\sigma}} -\frac{\epsilon^2}{c\sigma}, \\
	c_3 & = t -\frac{\epsilon^2}{c\sigma} \left(1 - e^{-\frac{t}{\epsilon^2/c\sigma}} \right).
	\end{align*}
	
	Note that due to the symmetry of angular integration,
	only the second term at the right-hand side of Eq.~\eqref{IeqFormula} will contribute to the macroscopic equilibrium flux,
	which leads to
	\begin{align}\label{FeqFormula}
	F^{eq}_{ij}
	=C_2 \int_{\mathcal{S}^2}
	\frac{c}{\epsilon}\vec{\Omega}\cdot\vec{n}_{ij}
	\left[ \frac{1}{4\pi} \frac{c}{\epsilon}\vec{\Omega} \cdot B_{g,\vec{x}}\left(\textbf{0},\vec{\Omega}\right)
	\right] \text{d}\vec{\Omega},
	\end{align}
	where
	\begin{align}\label{C2Formula}
	C_2 = \frac{1}{\Delta t} \int_{0}^{\Delta t} c_2 \text{d}t
	= \frac{2}{\Delta t} \left(\frac{\epsilon^2}{c\sigma}\right)^2 \left(1 - e^{-\frac{\Delta t}{\epsilon^2/c\sigma}}\right) - \frac{\epsilon^2}{c\sigma} - \frac{\epsilon^2}{c\sigma} e^{-\frac{\Delta t}{\epsilon^2/c\sigma}}.
	\end{align}
	
	For the non-equilibrium flux $F_{g,ij}^{fr}$,
	it indicates the numerical flux contributed by the initial specific radiation intensity during the free transport process.
	In the UGKP method, discrete particles are employed to represent the specific radiation intensity $I_g$,
	and the non-equilibrium flux is evaluated by tracking the particles.
	Specifically, the particle trajectory is fully resolved by
	\begin{equation}
	\vec{x} = \vec{x}^n + \frac{c}{\epsilon}\vec{\Omega} t_f,
	\end{equation}
	where $t_f$ denotes the free streaming time of the photon in the free transport process.
	It can be determined from the integral solution \eqref{integSolI},
	which can be re-written as
	\begin{equation*}
	I_g(\vec{x},t,\vec{\Omega})= \left(1-e^{-\frac{t-t_0}{\epsilon^2/c\sigma_g}}\right) B_g^{+}(\vec{x},t,\vec{\Omega})
	+e^{-\frac{t-t_0}{\epsilon^2/c\sigma_g}}I_{g,0}(\vec{x}-\frac{c}{\epsilon}\vec{\Omega}\left(t-t_0\right), \vec{\Omega}),
	\end{equation*}
	where $B_g^{+}$ is a near equilibrium distribution that can be expressed in an analytical way, i.e.,
	\begin{align*}
	B_g^{+}(\vec{x},t,\vec{\Omega})
	& = B_g(\vec{x},t,\vec{\Omega})
	+ \left(\frac{t e^{-\frac{t}{\epsilon^2/c\sigma_g}} }{1-e^{-\frac{t}{\epsilon^2/c\sigma_g}}} - \frac{\epsilon^2}{c \sigma_g}\right) \left[ \frac{c}{\epsilon}\vec{\Omega} \cdot B_{g,\vec{x}}(\vec{x},t,\vec{\Omega})
	+ B_{g,t}(\vec{x},t,\vec{\Omega}) \right].
	\end{align*}
	It can be found that the particles from $I_{g,0}$ has a probability of $e^{-\frac{t}{\epsilon^2/c\sigma_g}}$ to free transport
	and $(1-e^{-\frac{t}{\epsilon^2/c\sigma_g}})$ to be absorbed and re-emitted.
	Denoting the free streaming time before the first collision with the background material as $t_c$,
	the cumulative distribution function of $t_c$ gives
	\begin{equation}
	F\left(t_c < t\right) = 1 - e^{-\frac{t_c}{\epsilon^2/c\sigma_g}}.
	\end{equation}
	Then $t_c$ can be sampled by $t_c=-\frac{\epsilon^2}{c \sigma_g} \text{ln}\left(\eta\right)$ with a random number $\eta$ generated from a uniform distribution $U\left(0,1\right)$.
	Within one time step $\Delta t$,
	the free streaming time $t_f$ for particle $k$ in $g$-th group can be determined by
	\begin{equation}\label{particle tf formula}
	t_f = \min \left[-\frac{\epsilon^2}{c \sigma_g}\text{ln}\left(\eta\right), \Delta t\right].
	\end{equation}
	Thus, by counting the particles passing through cell interfaces,
	the overall non-equilibrium flux across all the interfaces of cell $i$, i.e.,
	$\sum\limits_{S_{ij}\in \partial V_i} -F_{g,ij}^{fr}S_{ij}\Delta t$
	can be calculated by
	\begin{equation}\label{FfrFomula}
	w_{g,i}^{fr} = \sum\limits_{k\in P_g\left(\partial \Omega_{i}^{+}\right)} \omega_k - \sum\limits_{k\in P_g\left(\partial \Omega_{i}^{-}\right)} \omega_k,
	\end{equation}
	where $P_g\left(\partial \Omega_{i}^{+}\right)$ is the particle set in $g$-th group moving into the cell $i$ during free transport process,
	$P_g\left(\partial \Omega_{i}^{-}\right)$ is the particle set in $g$-th group moving out of the cell $i$,
	and $\omega_k$ is radiation energy carried by particle $k$.
	
	As a result, with the equilibrium flux in Eq.~\eqref{FeqFormula} and non-equilibrium flux in Eq.~\eqref{FfrFomula},
	the cell averaged $\rho_{g}$ can be updated by
	\begin{equation*}
	\rho_{g,i}^{n+1} = \rho_{g,i}^n - \frac{\Delta t}{V_i} \sum\limits_{S_{ij}\in \partial V_i}F_{g,ij}^{eq}S_{ij}  + \frac{w_{g,i}^{fr}}{V_{i}} + \frac{c \sigma_{g,i}^{n+1} \Delta t}{\epsilon^2} \left( 4\pi B_{g,i}^{n+1} - \rho_{g,i}^{n+1} \right).
	\end{equation*}

	From Eq.~\eqref{particle tf formula},
	it can be found that all particles can be divided into two groups according to their free streaming time $t_f$:
	the collisionless particles ($t_f = \Delta t$) and the collisional particles ($t_f<\Delta t$).
	For the collisionless particles, their trajectories are fully tracked in the whole time step,
	and the time evolution of the collisionless particles is finished in the current step.
	While for the collisional particles,
	they would suffer encounter collisions with the background material being absorbed at time $t_f$.
	Therefore, these particles will be deleted after the free transport process,
	and the corresponding quantities carried by the collisional particles will be
	merged into the macroscopic flow variables of the relevant cell.
	The overall effect of the evolution of the photons emitted from the background material during the whole time step
	is modeled by the equilibrium flux $F_g^{eq}$ and the source term $B_g$.
	Therefore, from the updated specific radiation intensity $I_{g,i}^{n+1} = \frac{1}{4\pi}\rho_{g,i}^{n+1}$,
	the unsampled particles in cell $i$ can be directly obtained based on the conservation law
	\begin{equation}
	I^h_{g,i} = I^{n+1}_{g,i} - I^p_{g,i},
	\end{equation}
	where $I^p_{g,i}$ is the overall intensity of remaining collisionless particles in cell $i$ at the end of time step,
	i.e., $I_{g,i}^{p} = \frac{1}{4\pi}\frac{1}{V_i}\sum_{k\in V_{i}} \omega_k$.
	Besides, the macroscopic variable $I^h_{g, i}$ comes from the emitted particles that survive and remain in the domain after time evolution.
	It can be recovered by particle re-sampling,
	and the weight of each discrete particle can be determined by
	\begin{equation}
	\omega_k = \frac{4\pi I^h_g V}{N},
	\end{equation}
	where $V$ is the volume of the relevant cell,
	and $N$ is the number of particles to be sampled in this cell.
	
	The method described above is the so-called unified gas-kinetic particle (UGKP) method,
	where the specific intensity is fully represented by discrete particles.
	The UGKWP method is a further development of the UGKP method to achieve better efficiency
	by using a hybrid and adaptive representation of the specific intensity.
	From the UGKP method,
	it can be seen that the survived emission particles after time evolution are re-sampled from a given distribution.
	Theoretically, this part can be described by using an analytic way,
	and its contribution to non-equilibrium flux during the free transport process can be
	evaluated as same as those in the UGKS
	\begin{equation*}
	\begin{aligned}
	F^{fr,UGKS}_{g,ij}(I^h_{g,i})
	& = \frac{1}{\Delta t} \int_{0}^{\Delta t} \int_{\mathcal{S}^2} \frac{c}{\epsilon}\vec{\Omega}\cdot\vec{n}_{ij}
	\left[ e^{-\frac{t}{\epsilon^2/c\sigma_g}}I^h_{g,0}(-\frac{c}{\epsilon}t \vec{\Omega}, \vec{\Omega}) \right] \text{d}\vec{\Omega}\text{d}t                                         \\
	& = \int_{\mathcal{S}^2}
	\frac{c}{\epsilon}\vec{\Omega}\cdot\vec{n}_{ij} \left[ C_4 I_{g,0}^h\left(\vec{x},\vec{\Omega}\right)
	+ C_5 \frac{c}{\epsilon} \vec{\Omega}\cdot I_{g,0,\vec{x}}^h\left(\vec{x},\vec{\Omega}\right) \right]
	\text{d}\vec{\Omega},
	\end{aligned}
	\end{equation*}
	where the flux coefficients are
	\begin{equation*}
	\begin{aligned}
	C_4 & = \frac{1}{\Delta t} \frac{\epsilon^2}{c\sigma} \left(1 - e^{-\frac{\Delta t}{\epsilon^2/c\sigma}} \right) ,                                                                                     \\
	C_5 & = \frac{\epsilon^2}{c\sigma} e^{-\frac{\Delta t}{\epsilon^2/c\sigma}} - \frac{1}{\Delta t} \left(\frac{\epsilon^2}{c\sigma}\right)^2 \left(1 - e^{-\frac{\Delta t}{\epsilon^2/c\sigma}}\right).
	\end{aligned}
	\end{equation*}
	In order to preserve the capability to capture non-equilibrium physics induced by particles' free transport,
	the UGKWP method samples only the collisionless particles with the free transport time $t_f=\Delta t$ from $I^h_{g,i}$, i.e.,
	\begin{equation}\label{IhpFormula}
	I^{hp}_{g,i} = e^{-\frac{\Delta t}{\epsilon^2/c\sigma_g}} I^{h}_{g,i}.
	\end{equation}
	Then, the non-equilibrium  flux $w_{g, i}^{fr,p}$ contributed by the free streaming of the sampled collisionless particles can be calculated by particle tracking
	\begin{equation}\label{wfrpFormula}
	w_{g,i}^{fr,p} = \sum\limits_{k\in P_g\left(\partial \Omega_{i}^{+}\right)} \omega_k - \sum\limits_{k\in P_g\left(\partial \Omega_{i}^{-}\right)} \omega_k.
	\end{equation}
	Statistically, it corresponds to the free transport flux evaluated by the DVM
	\begin{equation*}
	\begin{aligned}
	F^{fr,DVM}_{g,ij}(I^{hp}_{g,i})
	& =e^{-\frac{\Delta t}{\epsilon^2/c\sigma_g}} \frac{1}{\Delta t} \int_{0}^{\Delta t} \int_{\mathcal{S}^2} \frac{c}{\epsilon}\vec{\Omega}\cdot\vec{n}_{ij}
	\left[ I_{g,0}^h\left(\vec{x},\vec{\Omega}\right) - \frac{c}{\epsilon} t\vec{\Omega}\cdot I_{g,0,\vec{x}}^h\left(\vec{x},\vec{\Omega}\right) \right] \text{d}\vec{\Omega}\text{d}t \\
	& =e^{-\frac{\Delta t}{\epsilon^2/c\sigma_g}}  \int_{\mathcal{S}^2}
	\frac{c}{\epsilon}\vec{\Omega}\cdot\vec{n}_{ij} \left[I_{g,0}^h\left(\vec{x},\vec{\Omega}\right)
	- \frac{\Delta t}{2} \frac{c}{\epsilon} \vec{\Omega}\cdot I_{g,0,\vec{x}}^h\left(\vec{x},\vec{\Omega}\right) \right]
	\text{d}\vec{\Omega}.
	\end{aligned}
	\end{equation*}
	Therefore, the unsampled particles represented by macroscopic variables, i.e., the so-called wave part,
	contribute to the non-equilibrium flux by
	\begin{equation*}
	\begin{aligned}
	F^{fr,wave}_{g,ij} &= F^{fr,UGKS}_{g,ij}(I^h_{g,i}) - F^{fr,DVM}_{g,ij}(I^{hp}_{g,i})  \\
	& =\int_{\mathcal{S}^2}
	\frac{c}{\epsilon}\vec{\Omega}\cdot\vec{n}_{ij} \left[ \left(C_4 - e^{-\frac{\Delta  t}{\epsilon^2/c\sigma_g}} \right) I_{g,0}^h\left(\vec{x},\vec{\Omega}\right)
	+ \left(C_5 + \frac{\Delta t}{2} e^{-\frac{\Delta t}{\epsilon^2/c\sigma_g}} \right) \frac{c}{\epsilon} \vec{\Omega}\cdot I_{g,0,\vec{x}}^h\left(\vec{x},\vec{\Omega}\right) \right]
	\text{d}\vec{\Omega}.
	\end{aligned}
	\end{equation*}

	Up to now, all the terms of fluxes in the UGKWP method have been determined,
	and thus the macroscopic radiation intensity $\rho$ can be updated by
	\begin{equation*}
	\rho_{g,i}^{n+1} = \rho_{g,i}^n - \frac{\Delta t}{V_i} \sum\limits_{S_{ij}\in \partial V_i}\left(F_{g,ij}^{eq} + F_{g,ij}^{fr,wave}\right)S_{ij}  + \frac{w_{g,i}^{fr,p}}{V_{i}} + \frac{c \sigma_{g,i}^{n+1} \Delta t}{\epsilon^2} \left( 4\pi B_{g,i}^{n+1} - \rho_{g,i}^{n+1} \right).
	\end{equation*}
	Accordingly, the implicitly discretized RTE system Eq.~\eqref{macroEquImplicit} can be re-written as
	\begin{equation}\label{macroSysFinal}
	\begin{cases}
	\rho_{g,i}^{n+1} = \rho_{g,i}^n - \frac{\Delta t}{V_i} \sum\limits_{S_{ij}\in \partial V_i}\left(F_{g,ij}^{eq} + F_{g,ij}^{fr,wave}\right)S_{ij}  + \frac{w_{g,i}^{fr,p}}{V_{i}} + \frac{c \Delta t}{\epsilon^2} \left(\sigma_{g,i}^{n+1} 4\pi B_{g,i}^{n+1} - \sigma_{g,i}^{n+1} \rho_{g,i}^{n+1} \right), \\
	C_vT_i^{n+1} = C_vT_i^{n} + \sum\limits_{g=1}^{G} \frac{\Delta t}{\epsilon^2}\left(\sigma_{g,i}^{n+1} \rho_{g,i}^{n+1} - \sigma_{g,i}^{n+1} 4\pi B_{g,i}^{n+1} \right).
	\end{cases}
	\end{equation}
	Similar to the treatment in \cite{UGKS-radiative-sun2015asymptotic, rad-wp-implicit-liu2023implicit},
	the implicit system of Eq.~\eqref{macroSysFinal} will be solved by the source iteration method,
	and the details will be provided in Appendix B.
	Finally, the algorithm of the multi-scale UGKWP method for the multi-frequency radiation transport system can be summarized below.
	\begin{algorithm}[htb]
		\caption{Multi-scale UGKWP method for multi-frequency radiation transport system.}
		\begin{algorithmic}[1]
			\STATE Start with $t=0$
			\WHILE{$t < t_{end}$}
			\STATE Determine $t_f$ based on Eq.~\eqref{particle tf formula} for the remaining particles (if existing);
			\STATE Sample the free transport particles with $t_f = \Delta t$ according to Eq.~\eqref{IhpFormula};
			\STATE Stream the existing particles with $t_f$, and get the flux $w_{g,i}^{fr,p}$ based on Eq.~\eqref{wfrpFormula};
			\STATE Update macroscopic variables $\rho_{g,i}^{n+1}$, $g=1,...,G$, $T_i^{n+1}$ in Eq.~\eqref{macroSysFinal} by source iteration;
			\STATE Absorb the particles with $t_f < \Delta t$.
			\STATE $t = t + \Delta t$
			\ENDWHILE	
			\STATE End
		\end{algorithmic}
	\end{algorithm}
	
\begin{figure}[htbp]
		\centering
		\subfigure{
			\includegraphics[height=4.5cm]{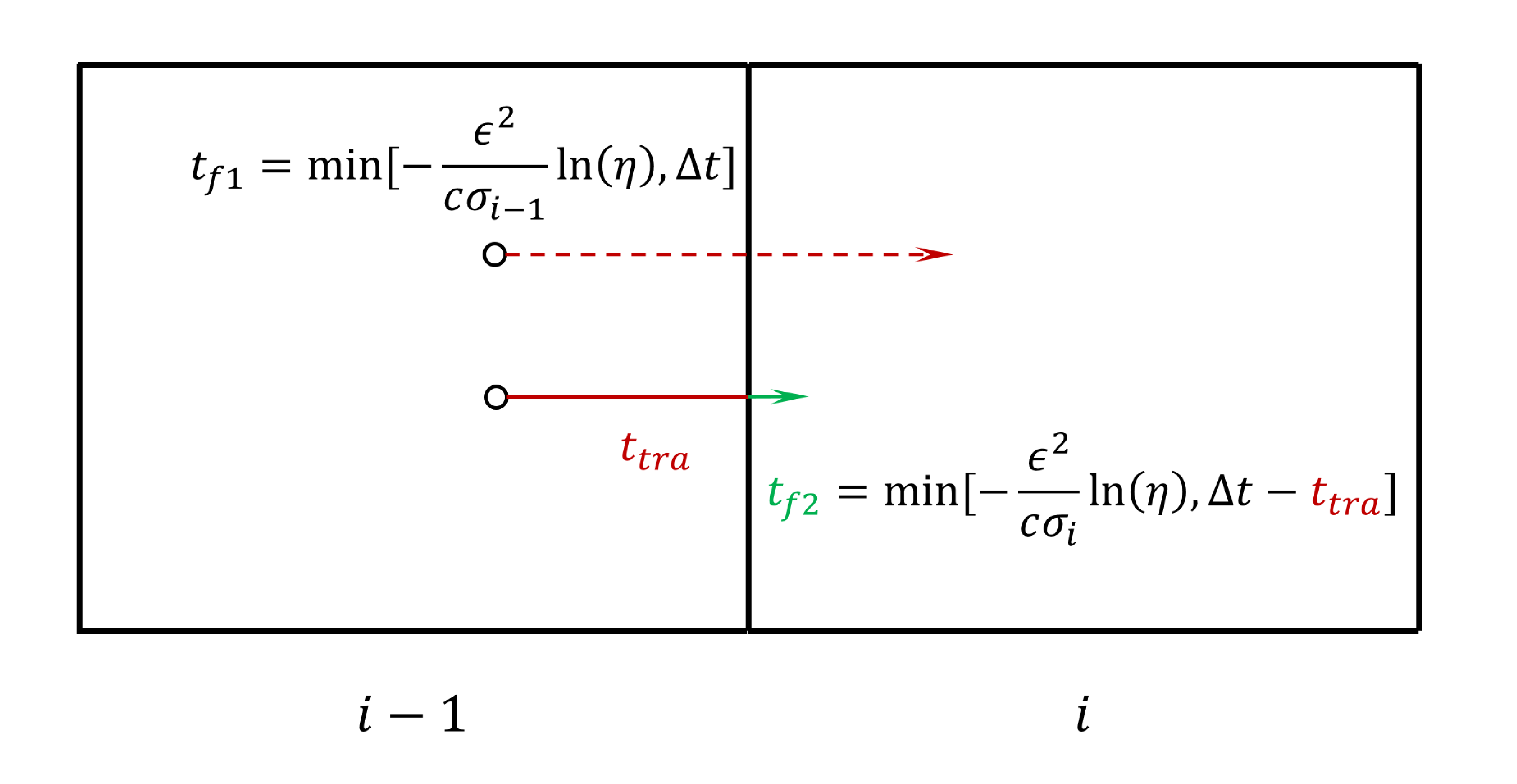}}
		\caption{The schematic diagram of tracking procedure of particles crossing one interface.}
		\label{sketch particle sample}
	\end{figure}

	\begin{rmk}
		In the radiation transfer system, the values of $\sigma$ in two adjacent cells may be dramatically different.
		In order to improve the spatial resolution of local physics,
		the free streaming time $t_f$ of a particle needs to be reset when it travels through a cell interface and moves into a new cell.
	\end{rmk}
	As illustrated in Fig.~\ref{sketch particle sample},
	considering the case that a particle moving into cell $i$ from cell $i-1$,
	the secondary free streaming time $t_{f2}$ will be determined by the local state in the new cell $i$
	\begin{equation}
	t_{f2} = \min \left[-\frac{\epsilon^2}{c \sigma_i}\text{ln}\left(\eta\right), \Delta t - t_{tra}\right],
	\end{equation}
	where $t_{tra}$ stands for the time that the particle has already been tracked in cell $i-1$,
	and $\sigma_i$ is the opacity of cell $i$.
	
	\begin{rmk}
		For the multi-frequency RTE system, particles should carry different frequencies to recover the photon's frequency spectra.
		For better consistency with the reality, the frequency of each particle is sampled by following the distribution of $B\left(\nu\right)$,
		instead of taking a fixed value for all the particles inside the $g$-th frequency interval, such as the mid-value $\nu_{g, mid}$.
	\end{rmk}
	
	In this study, for a balance of computational efficiency and accuracy,
	linear distribution of $B_g\left(\nu\right)$ is assumed inside the $g$-th frequency interval
	\begin{equation}\label{Blinear}
	B_g\left(\nu\right) = B\left(\nu_{g,mid}\right) + k_g \left(\nu - \nu_{g,mid}\right),
	\end{equation}
	with
	\begin{equation}\label{kg}
	k_g = \frac{B\left(\nu_{g+1/2}\right) - B\left(\nu_{g-1/2}\right)}{\nu_{g+1/2} - \nu_{g-1/2}}.	
	\end{equation}
	With the linear distribution of $B_g\left(\nu\right)$, the frequency value carried by sampled particles can be determined easily, and the details are introduced in Appendix A.

	\begin{rmk}
		$F^{fr, wave}_{ij}$ stands for the free-transport flux contributed from the un-sampled particles.
		If the collisionless particles were not indeed sampled, e.g.,
		due to a small value of $e^{-\frac{\Delta t}{\epsilon^2/c\sigma}}$,
		the free-transport flux should regress to the same as in UGKS, i.e., $F^{fr, wave}_{ij} = F^{fr, UGKS}_{ij}$.
	\end{rmk}
	
	\section{Numerical results}
	In this section, a series of benchmark problems covering different flow regimes will be calculated to validate the proposed method.
	In all cases, the unit of length is centimeter ${\rm cm}$,
	the mass unit is gram ${\rm g}$,
	the time unit is nanosecond ${\rm ns}$,
	the temperature unit is kilo electronvolt ${\rm keV}$,
	and the energy unit is $10^9$ Joules ${\rm GJ}$.
	Thus, the speed of light is $c=29.98{\rm cm/ns}$,
	and the radiation constant is $a=0.01372 {\rm GJ/\left(cm^3\cdot keV^4\right)}$.
	The time step is determined by $\Delta t = \text{CFL} \times \frac{\Delta x}{c}$ with $\text{CFL}=1.0$.
	Besides, for the frequency-dependent cases,
	the frequency space $\nu$ is logarithmically discretized into $G=24$ groups.
	Theoretically we have $h\nu_{min}=0$ and $h\nu_{max}= \infty $,
	but in this paper we take $h\nu_{min}=10^{-2} {\rm keV}$ and $h\nu_{max}=10^{2} {\rm keV}$.
	For the discrete $g$-th group, the Simpson's rule, based on $\nu_{g-1/2}$, $\nu_{g, mid}$, and $\nu_{g+1/2}$,
	is employed to calculate the overall opacity,
	which will be used to determine the flow regime for $g$-th group photons.
	For all groups, the particle weight $\omega$ is taken as $\omega = 10^{-4}V$ unless further special notification.

	\subsection{Homogeneous Marshak wave problem}
\begin{figure}[htbp]
		\centering
		\subfigure{
			\includegraphics[height=6.5cm]{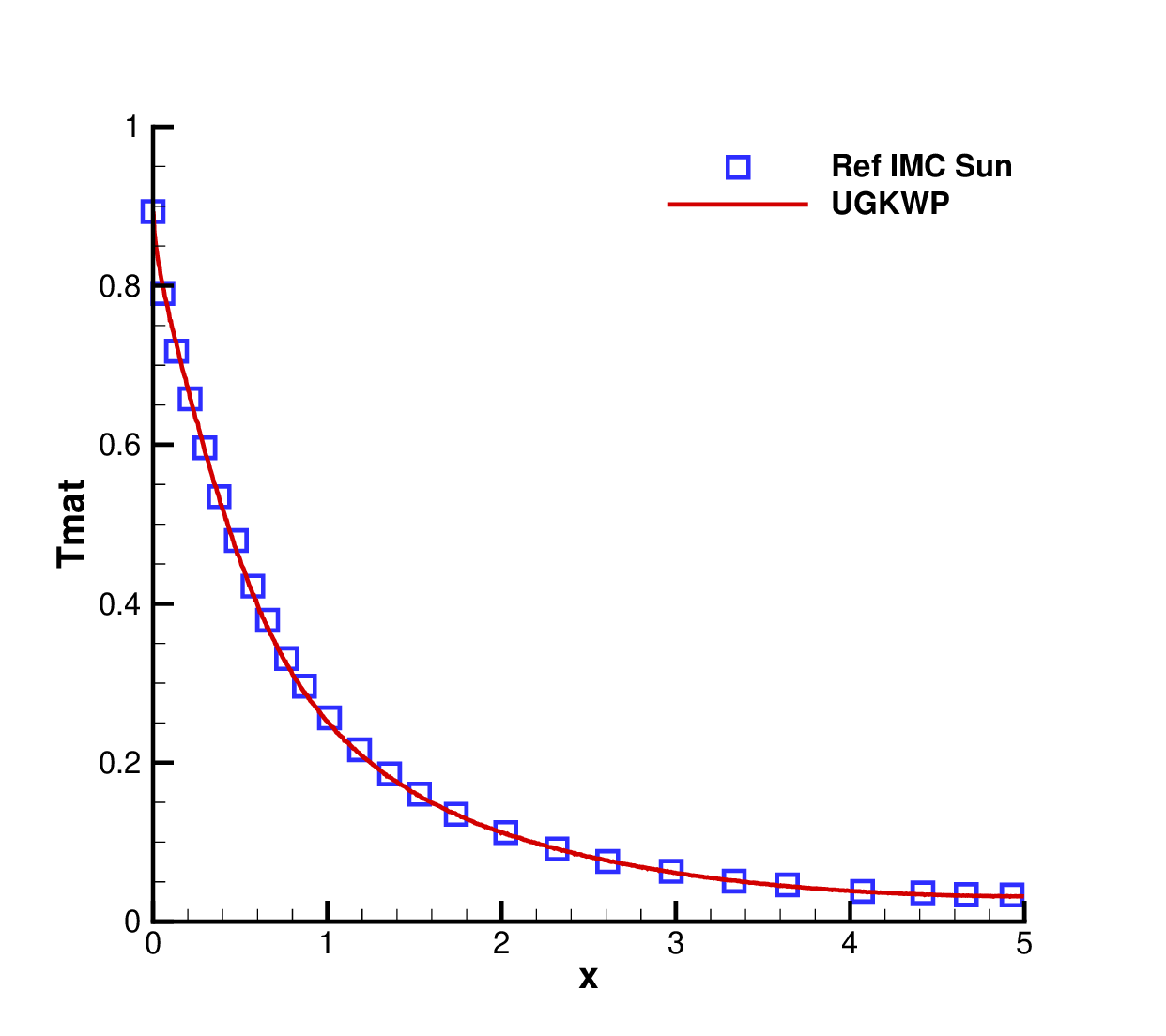}}
		\quad
		\subfigure{
			\includegraphics[height=6.5cm]{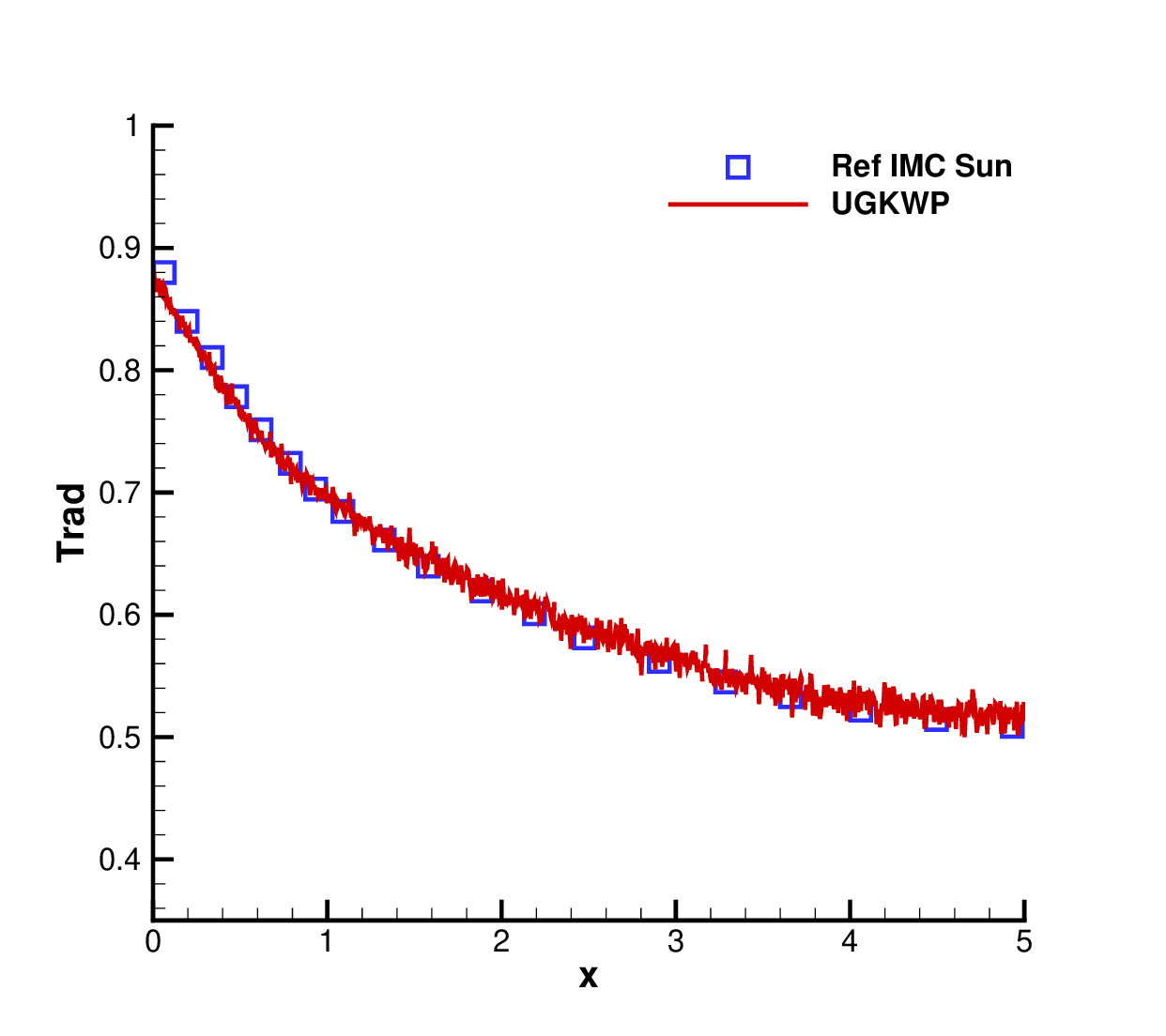}}
		\caption{Material temperature and radiation temperature of homogeneous Marshak wave problem with $\sigma_0 = 10 {\rm keV^{7/2} / cm}$ at $t=1 {\rm ns}$.}
		\label{Marshark wave 10}
	\end{figure}

\begin{figure}[htbp]
		\centering
		\subfigure{
			\includegraphics[height=6.5cm]{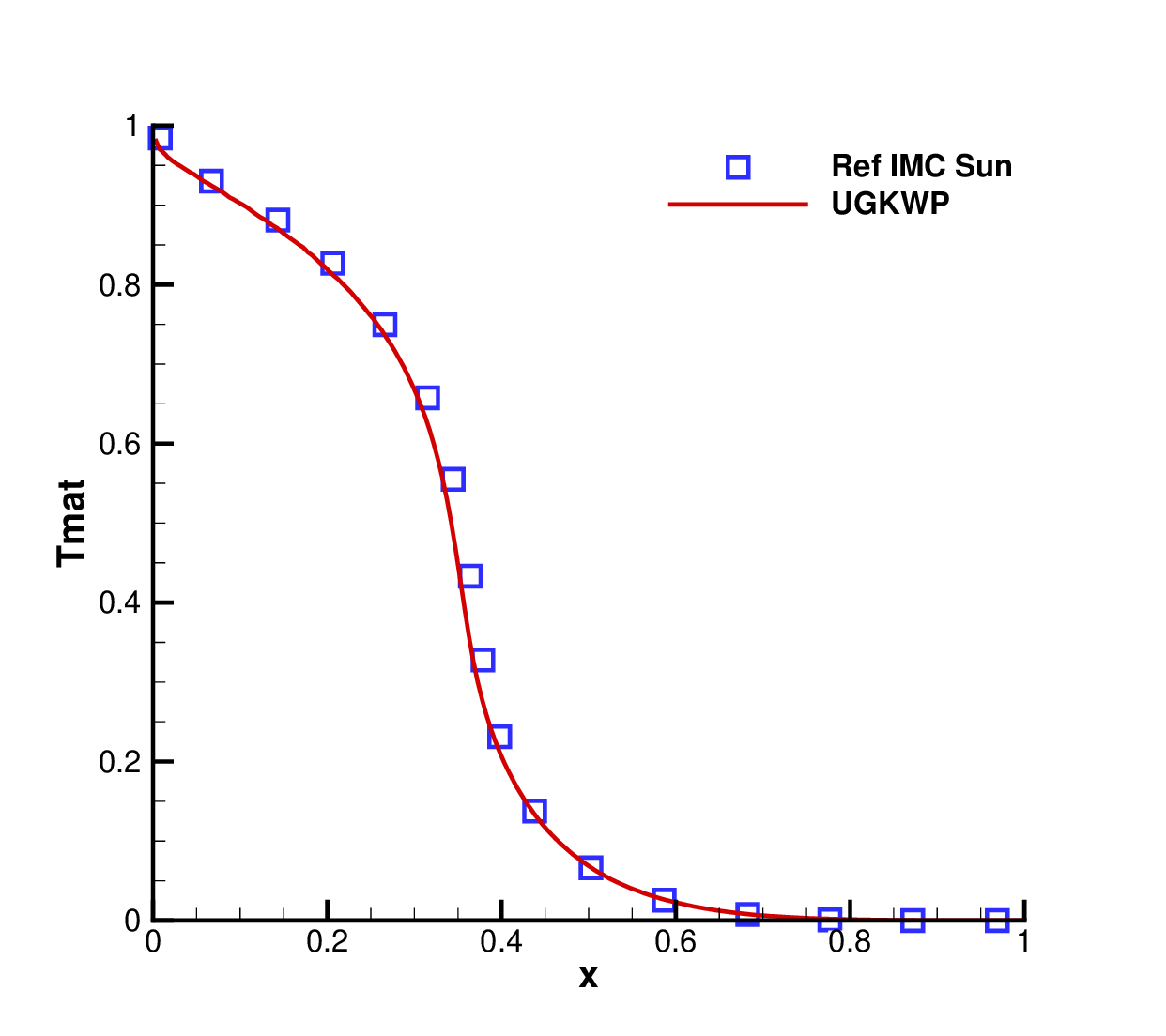}}
		\quad
		\subfigure{
			\includegraphics[height=6.5cm]{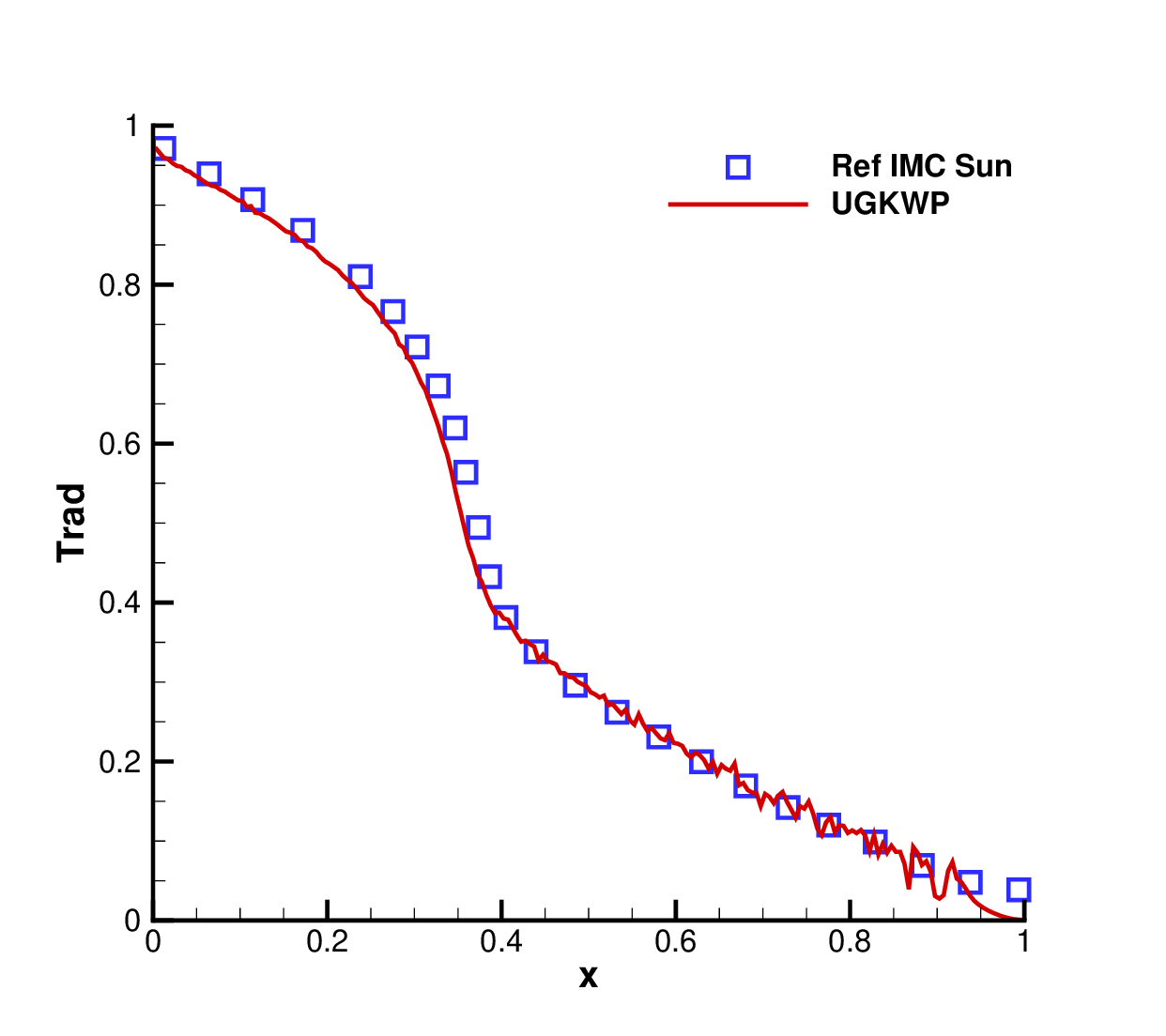}}
		\caption{Material temperature and radiation temperature of homogeneous Marshak wave problem with $\sigma_0 = 1000 {\rm keV^{7/2} / cm}$ at $t=1 {\rm ns}$. Note that only the results in $\left[0 {\rm cm}, 1 {\rm cm}\right]$ are shown.}
		\label{Marshark wave 1000}
	\end{figure}

	The first case is the homogeneous frequency-dependent Marshak wave problem.
	In this example, the heat capacity is a constant $C_v=0.1 {\rm GJ/\left(cm^3 \cdot keV\right)}$,
	and the opacity depends on both the temperature of background material $T$ and the photon's frequency $\nu$,
	which can be written in the following form,
	\begin{equation}\label{sigmaMarsharkMF}
	\sigma\left(\nu, T\right) = \frac{\sigma_0}{\left(h\nu\right)^3 \sqrt{kT}}.
	\end{equation}
	The computational domain is $\left[0{\rm cm},5{\rm cm}\right]$ with uniform cell size $\Delta x = 5\times10^{-3}{\rm cm}$.
	In the whole domain, the initial condition is $T=10^{-3} {\rm keV}$,
	and the radiation intensity follows Planck's distribution with radiation temperature $T_r=10^{-3} {\rm keV}$.
	At the left boundary, the incident intensity with $T_r=1.0 {\rm keV}$ also follows Planck's distribution;
	while the reflective boundary condition is employed for the right boundary.
	Both the thinner case with $\sigma_0 = 10 {\rm keV^{7/2}/cm}$,
	and the thicker case with $\sigma_0 = 1000 {\rm keV^{7/2}/cm}$ are calculated by UGKWP method,
	and the results at $t=1 {\rm ns}$ are presented in Figure~\ref{Marshark wave 10} and Figure~\ref{Marshark wave 1000}, respectively.
	The results agree well with the solution of IMC in both optically thin and thick regimes \cite{rad-mf-UGKS-sun2015asymptotic}.
	Note that in the thicker case with $\sigma_0 = 1000 {\rm keV^{7/2}/cm}$ the particle weight $\omega$ is taken as $\omega = 10^{-5}V$ to obtain the results with lower noise, and only the domain of $[0{\rm cm}, 1{\rm cm}]$ is shown here.
	
	\subsection{Heterogeneous Marshak wave problem}
	Another example is the heterogeneous Marshak wave problem,
	where the opacity is different in space.
	As a result, there exists the zone of thin-thick-joint,
	where the opacity dramatically varies within two adjacent cells,
	which is very challenging for a numerical scheme.
	In this example, the opacity $\sigma$ is calculated by Eq.~\eqref{sigmaMarsharkMF} as well.
	Two cases under different conditions are tested.
	
\begin{figure}[htbp]
		\centering
		\subfigure{
			\includegraphics[height=6.5cm]{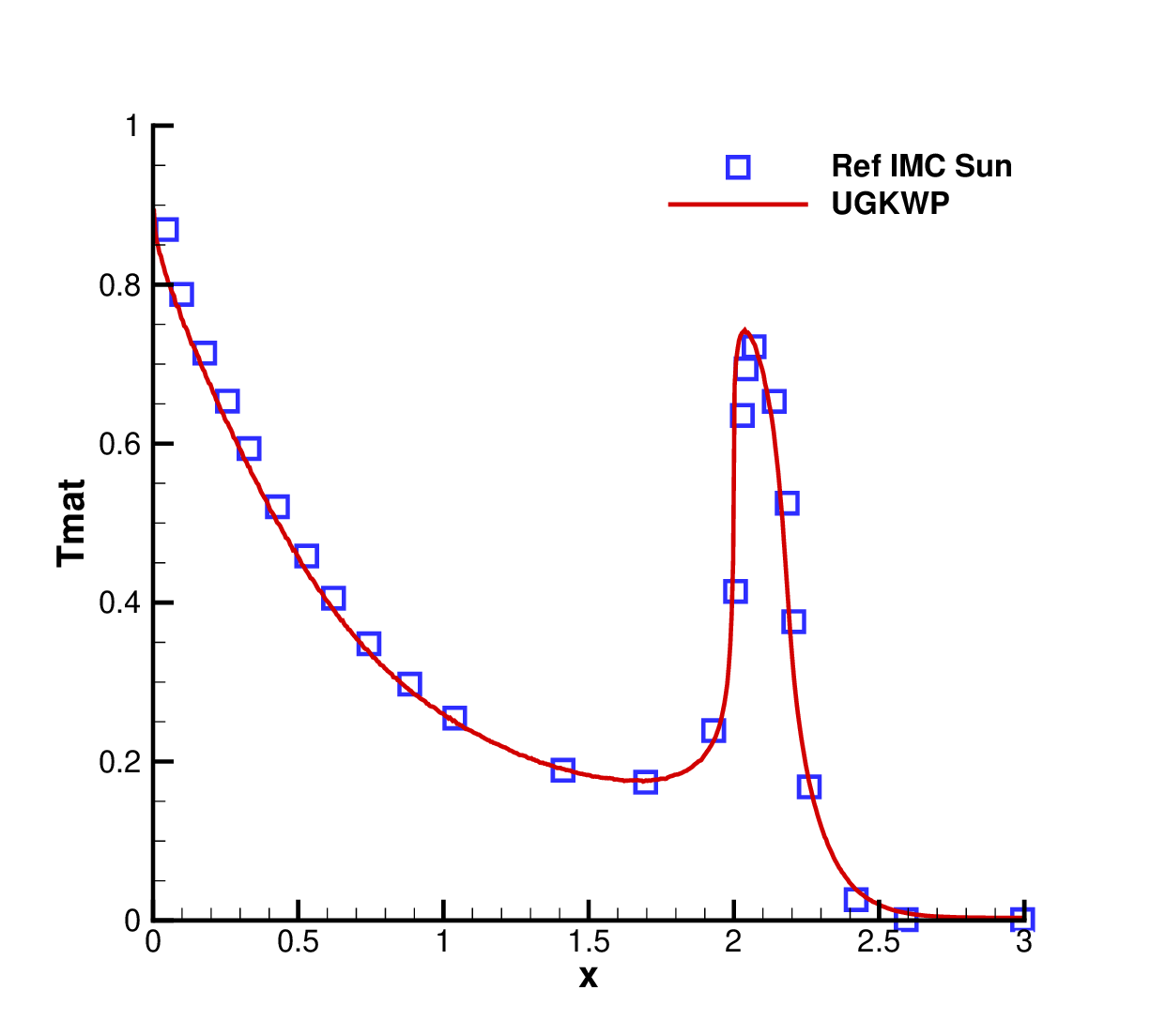}}
		\quad
		\subfigure{
			\includegraphics[height=6.5cm]{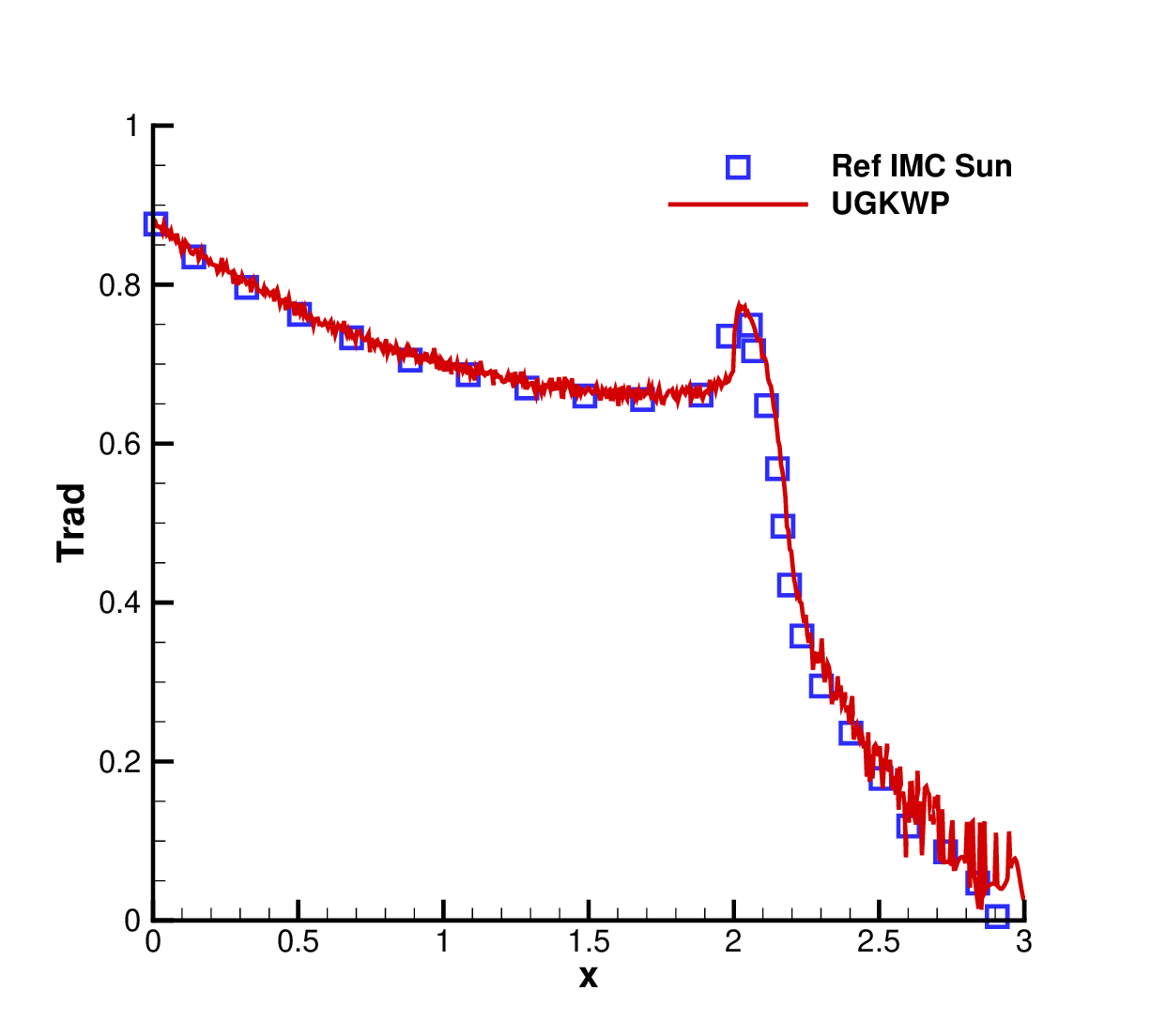}}
		\caption{Material temperature and radiation temperature of the heterogeneous thin-to-thick case at $t=1.0{\rm ns}$.}
		\label{Marshark wave hetecase2}
	\end{figure}

	For the first case, the computational domain is $\left[0{\rm cm},3{\rm cm}\right]$ with the uniform cell size $\Delta x = 5\times10^{-3}{\rm cm}$.
	$\sigma_0$ is spatially heterogeneous in the computational domain, which has
	\begin{equation*}
	\sigma_0=
	\left\{\begin{aligned}
	& 10 ~{\rm keV^{7/2}/cm},   & 0{\rm cm} \le x \le 2{\rm cm}, \\
	& 1000 ~{\rm keV^{7/2}/cm}, & 2{\rm cm} < x \le 3{\rm cm}.
	\end{aligned}\right.
	\end{equation*}
	The initial and boundary conditions are the same as the above homogeneous Marshak wave problem.
	As a result, there is a thin-to-thick joint cell at $x=2{\rm cm}$.
	Figure~\ref{Marshark wave hetecase2} presents the distribution of material and radiation temperature at $t=1{\rm ns}$,
	which agrees well with the IMC solution \cite{rad-mf-UGKS-sun2015asymptotic}.
	
\begin{figure}[htbp]
		\centering
		\subfigure{
			\includegraphics[height=6.5cm]{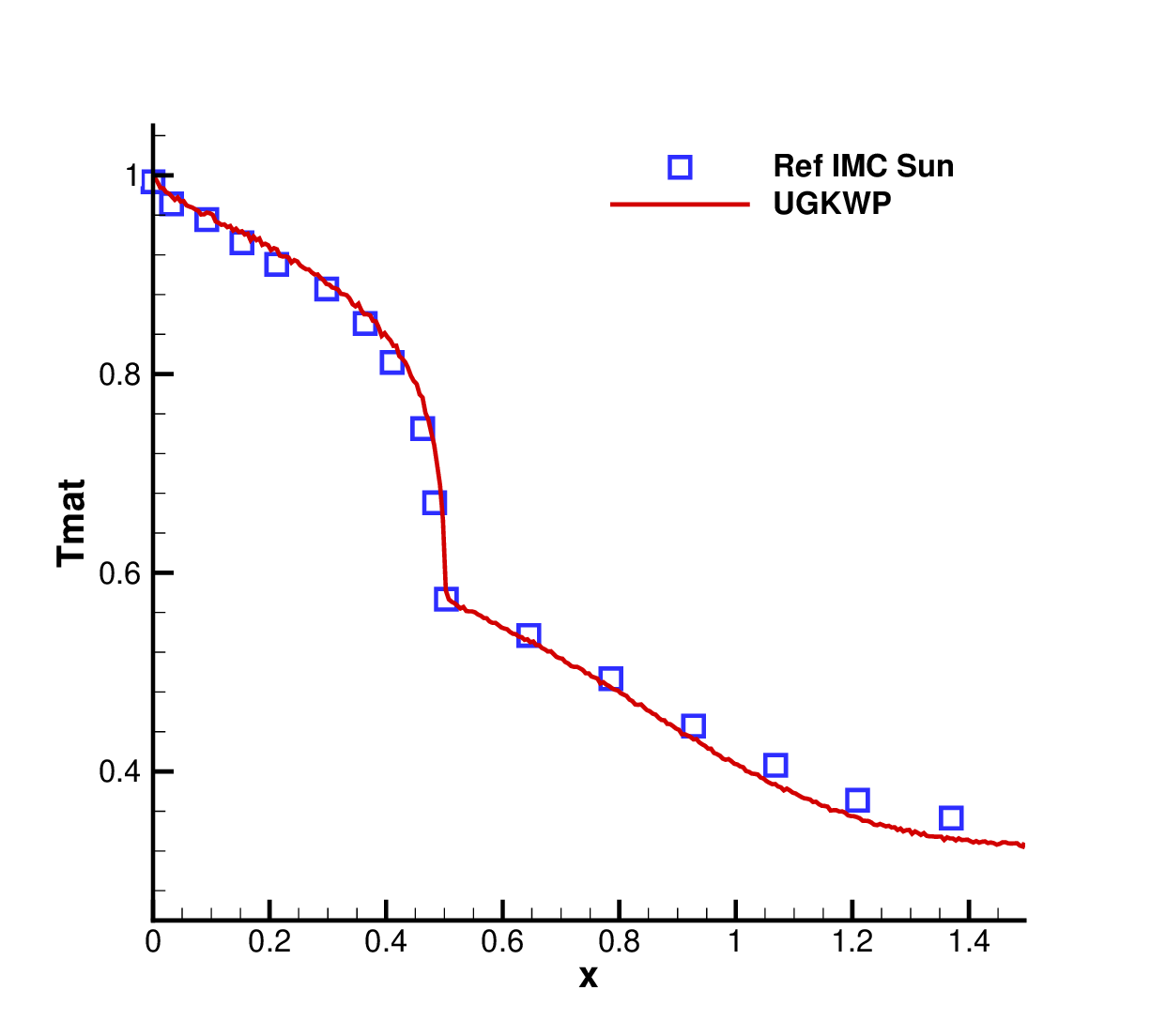}}
		\quad
		\subfigure{
			\includegraphics[height=6.5cm]{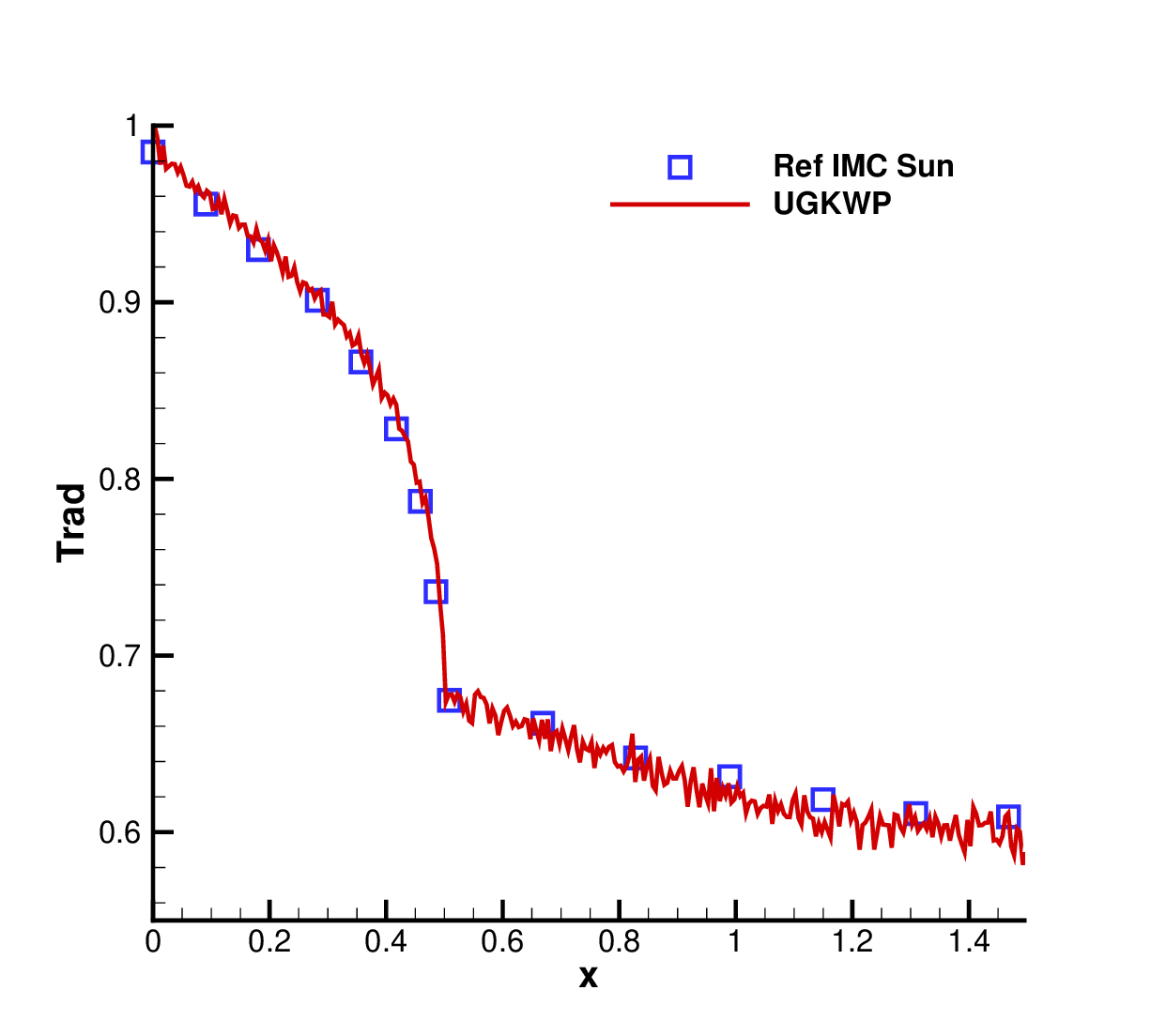}}
		\caption{Material temperature and radiation temperature of the heterogeneous thick-to-thin case at $t=5.0{\rm ns}$.}
		\label{Marshark wave hetecase3}
	\end{figure}

	For the second case, the computation domain is $\left[0{\rm cm}, 1.5{\rm cm}\right]$,
	and the same spatial resolution is adopted as that in the first case.
	The heterogeneous $\sigma_0$ is taken as
	\begin{equation*}
	\sigma_0=
	\left\{\begin{aligned}
	& 1000 ~{\rm keV^{7/2}/cm}, & 0{\rm cm} \le x \le 0.5{\rm cm}, \\
	& 10 ~{\rm keV^{7/2}/cm},   & 0.5{\rm cm} < x \le 1.5{\rm cm},
	\end{aligned}\right.
	\end{equation*}
	which leads to the thick-to-thin joint cell.
	In this case, the numerical collision time $\tau_{num}$ is employed to increase the sampled particles in the cells of thin-thick-joint,
	which means $e^{-\Delta t / \tau_{num}}$ will be used in Eq.~\eqref{IhpFormula}.
	Specifically, in this paper the $\tau_{num}$ is calculated by
	\begin{equation*}\label{tauNum}
	\tau_{num} = \frac{\epsilon^2}{c\sigma} + C \times \frac{T^{l} - T^{r}}{T^{l} + T^{r}} \Delta t,
	\end{equation*}
	where $T^l, T^r$ are the material temperatures on the left and right sides of a cell interface,
	and $C = 10^5$ is used in this case.
	The results at $t=5{\rm ns}$ shown in Figure~\ref{Marshark wave hetecase3} indicate that UGKWP performs well and can recover the IMC solution \cite{rad-mf-UGKS-sun2015asymptotic}.
	
	\subsection{Larsen's test problem}
	The Larsen problem is also tested in this study.
	For this case, the heat capacity is taken as $C_v=0.05109 {\rm GJ/\left(cm^3 \cdot keV\right)}$,
	and the photo-ionization absorption is modeled in the formula of opacity by
	\begin{equation*}
	\sigma =\gamma_0 \frac{1-e^{-h\nu/kT}}{\left(h\nu\right)^3},
	\end{equation*}
	where the distribution of $\gamma_0$ is heterogeneous in space with,
	\begin{equation*}
	\gamma_0=
	\left\{\begin{aligned}
	& 1 ~{\rm keV^{3}/cm},    & 0{\rm cm} \le x \le 2{\rm cm}, \\
	& 1000 ~{\rm keV^{3}/cm}, & 2{\rm cm} < x \le 3{\rm cm},   \\
	& 1 ~{\rm keV^{3}/cm},    & 3{\rm cm} < x \le 4{\rm cm}.
	\end{aligned}\right.
	\end{equation*}
	As shown above, the opacity has a dramatic spatial variation with three order-of-magnitude differences.
	As a result, the flow regime may change from the limiting of photon's free transport to the limiting of thermal diffusion within two adjacent cells.
	
\begin{figure}[htbp]
		\centering
		\subfigure{
			\includegraphics[height=6.5cm]{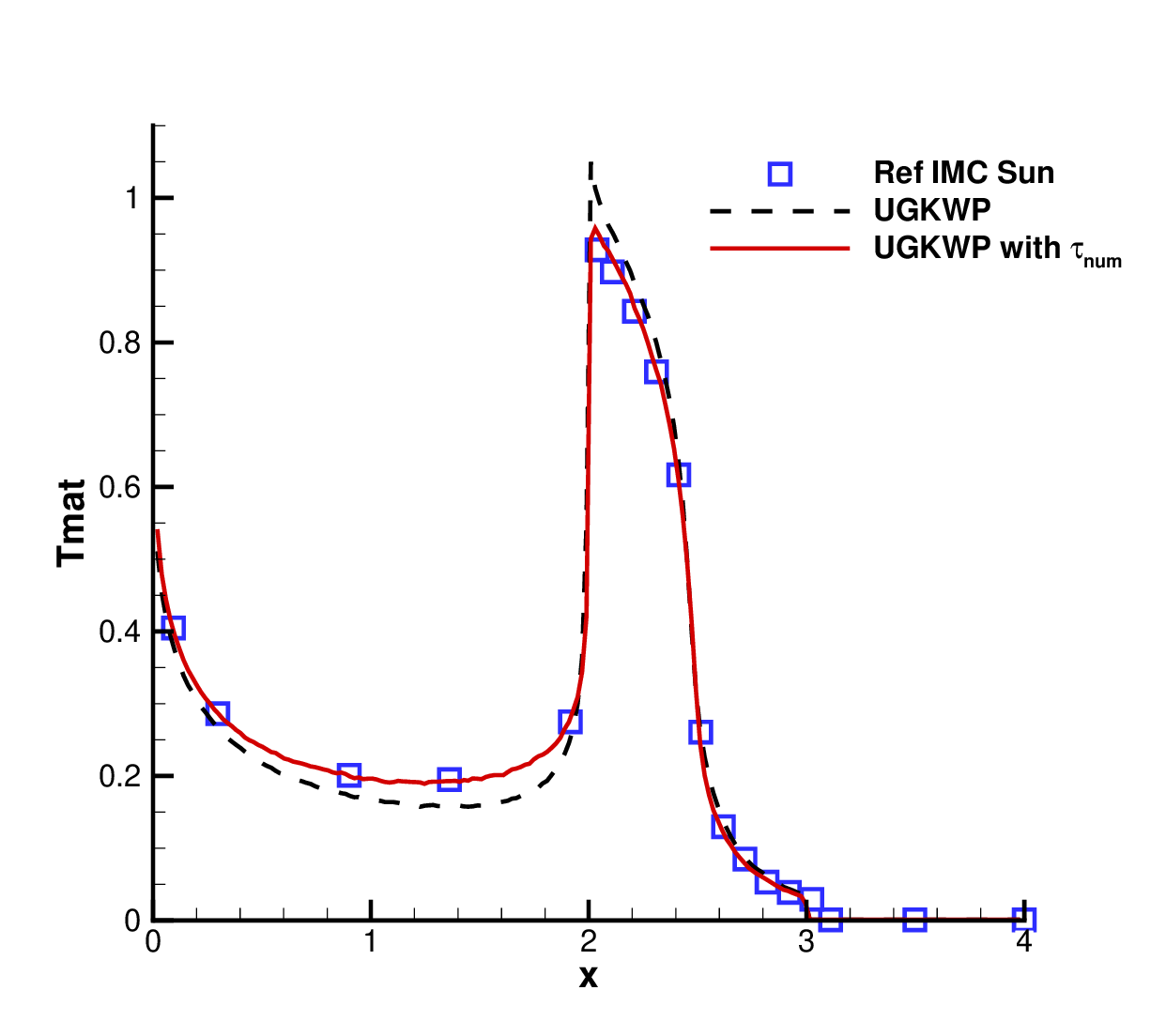}}
		\quad
		\subfigure{
			\includegraphics[height=6.5cm]{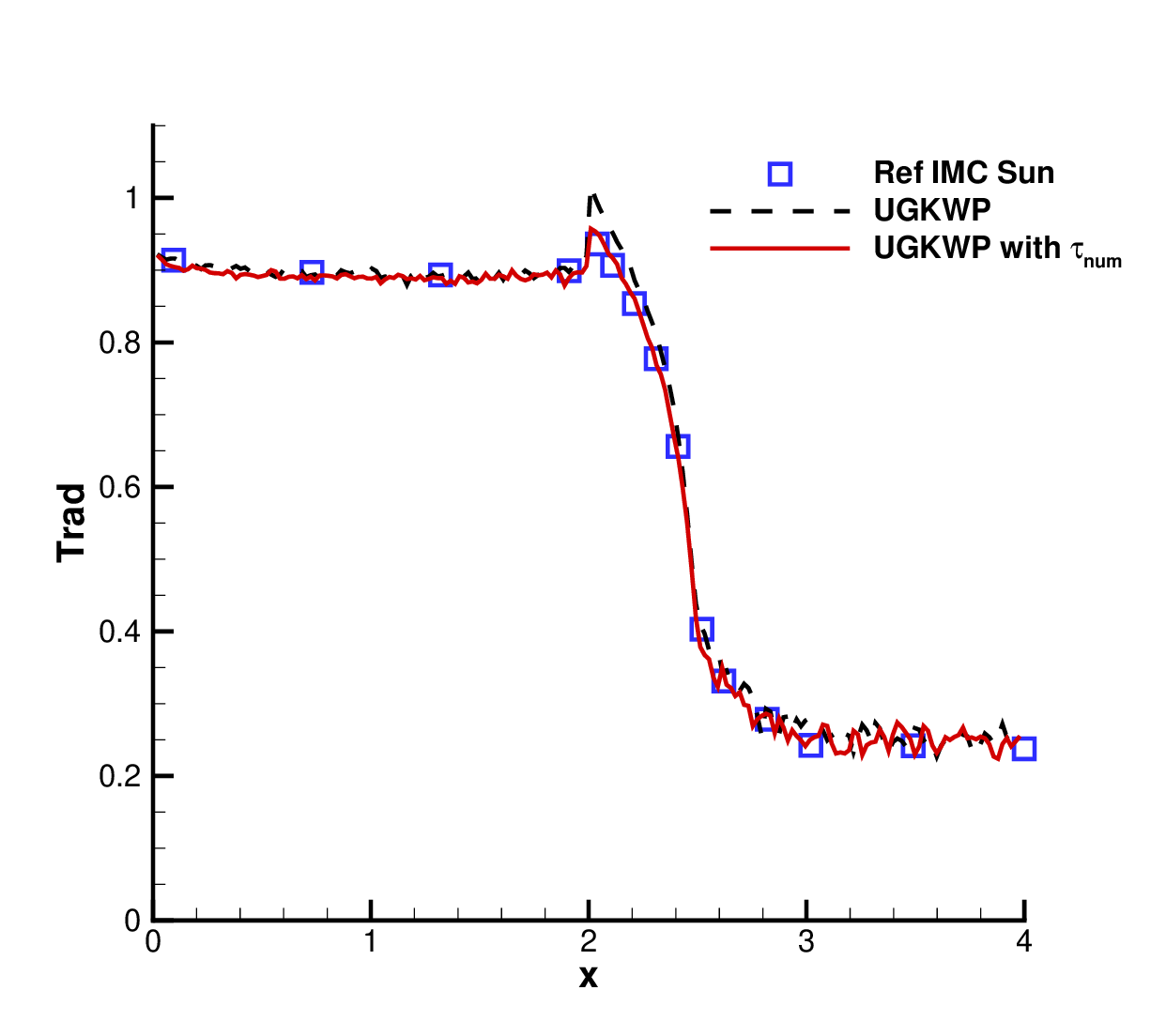}}
		\caption{Material temperature and radiation temperature of Larsen problem at $t=0.9{\rm ns}$.}
		\label{LarsenCase}
	\end{figure}

	The computational domain is $\left[0{\rm cm},4{\rm cm}\right]$ with uniform cell size $\Delta x = 2\times10^{-2}{\rm cm}$.
	The initial condition is in equilibrium state with $T=10^{-3}{\rm keV}$.
	At the left boundary, the incident intensity with $T_r=1.0{\rm keV}$ also follows Planck's distribution;
	and the free boundary condition is employed for the right boundary.
	In this case, the $\tau_{num}$ in Eq.~\eqref{tauNum} with $C=10^3$ is also employed for sampling particles in the cells of thin-thick-joint.
	The distribution of material temperature and radiation temperature at $t=0.9{\rm ns}$ are shown in Figure~\ref{LarsenCase}.
	The results indicate that, with the modification by $\tau_{num}$ in sampling particles,
	the numerical oscillations in the zone of thin-thick-joint, around $x=2{\rm cm}$,
	can be effectively avoided, and the profiles by the UGKWP method agree well with the IMC solution \cite{rad-mf-UGKS-sun2015asymptotic}.
	
	\subsection{Olson 1D case}
\begin{figure}[htbp]
		\centering
		\subfigure{
			\includegraphics[height=6.5cm]{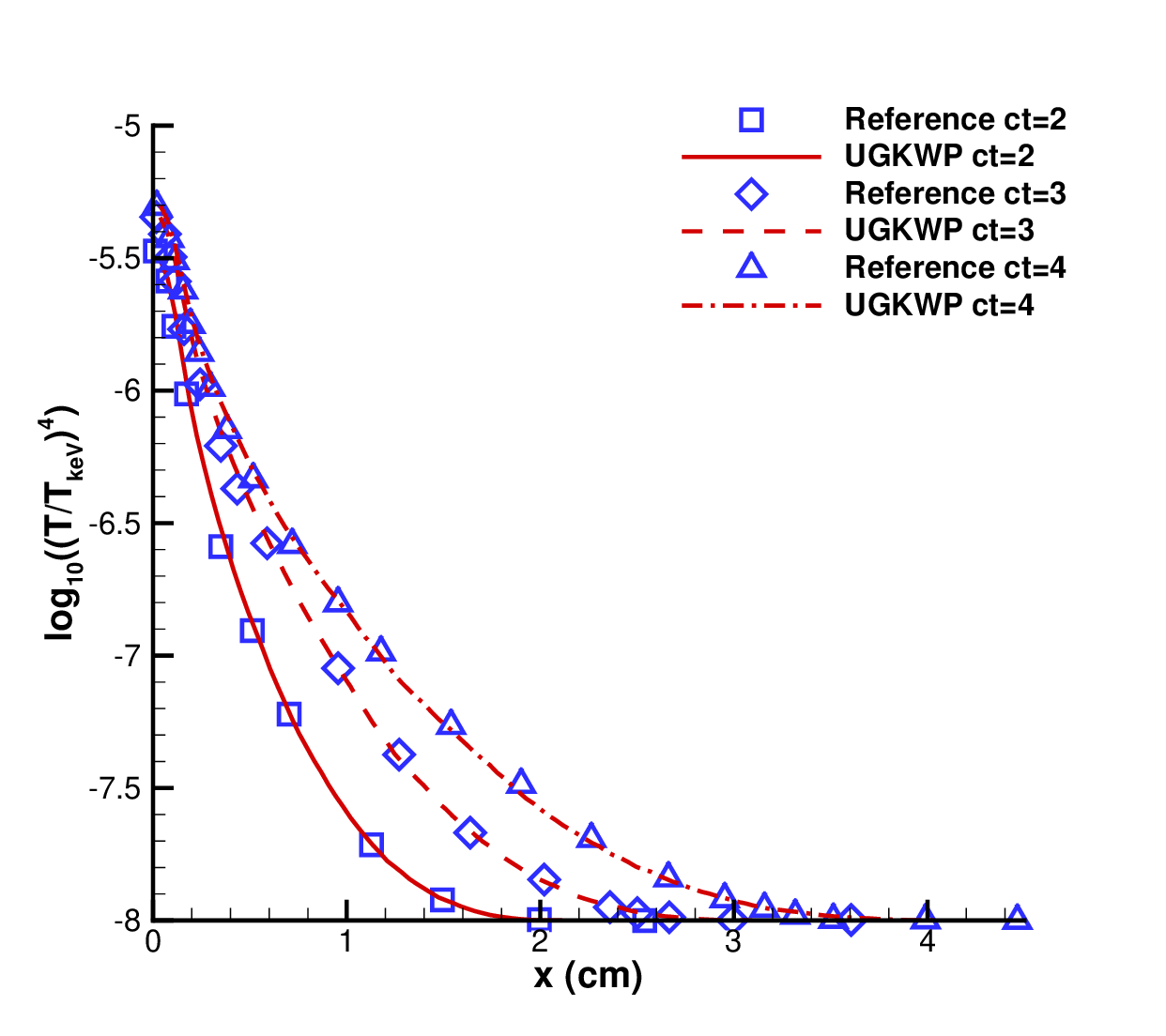}}
		\quad
		\subfigure{
			\includegraphics[height=6.5cm]{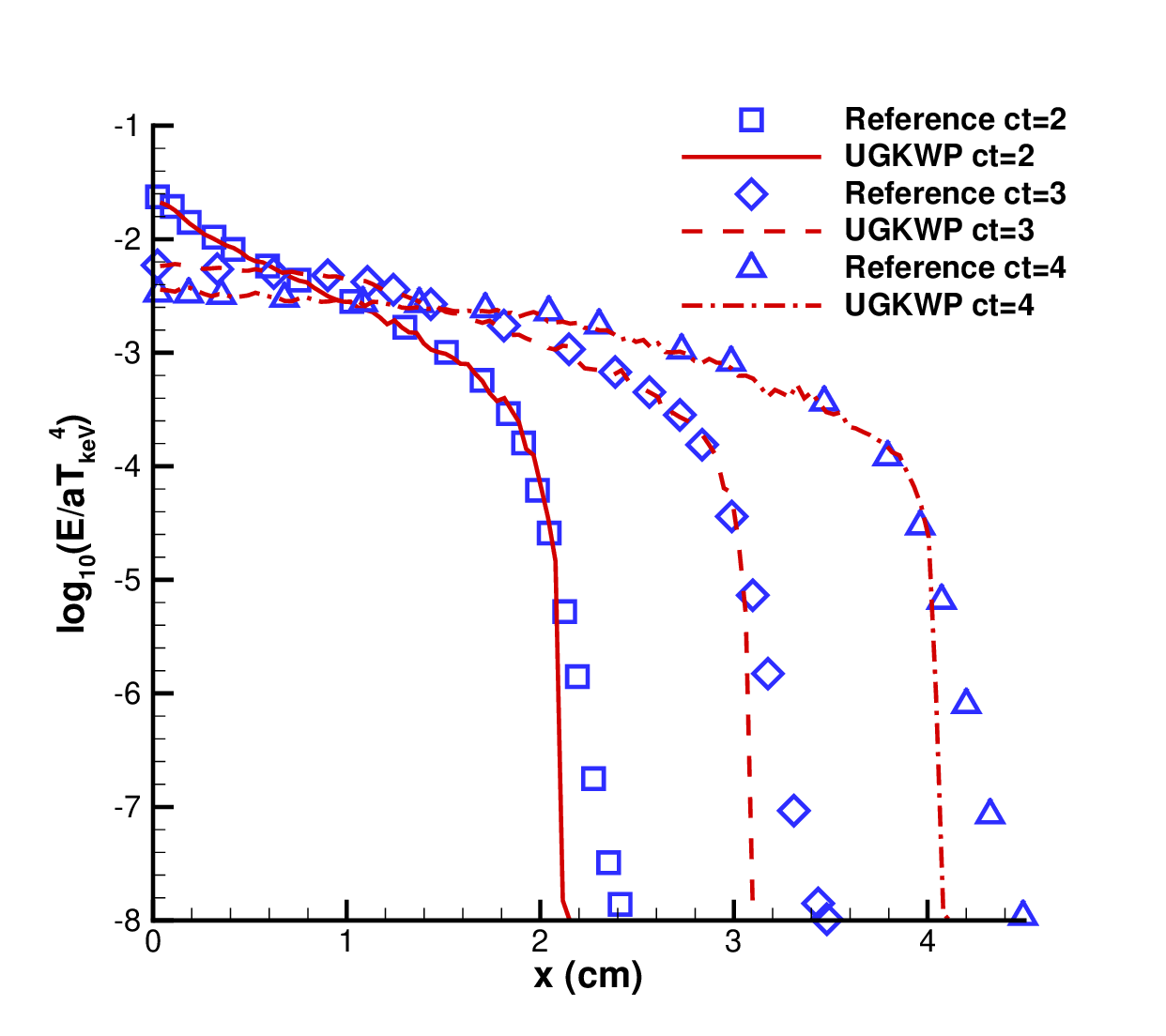}}
		\caption{Material temperature and radiation intensity of Olson 1D problem at $ct=2, 3, 4$.}
		\label{Olson1D}
	\end{figure}

	Further, the Olson 1D case which involves the photon's transport in carbon-hydrogen foam is tested by the frequency-dependent UGKWP method. The background material in the domain is carbon-hydrogen foam, of which the opacity depends on the photon's frequency and material temperature. Particularly, the opacity can be calculated by $\sigma = \rho \kappa$, where,
	\begin{equation*}
	\kappa=
	\begin{cases}
	\min\left(10^7, 10^9 \left(T/T_{keV}\right)^2\right), & h\nu \le 0.008 {\rm keV},                \\
	\frac{3 \times10^6 \left(\frac{0.008{\rm keV}}{h\nu}\right)^2}{1+200 \left(T/T_{keV}\right)^{1.5}},
	& 0.008 {\rm keV} < h\nu \le 0.3{\rm keV}, \\
	\frac{3 \times10^6 \left(\frac{0.008{\rm keV}}{h\nu}\right)^2 \left(\frac{0.3{\rm keV}}{h\nu}\right)^{0.5}}{1+200 \left(T/T_{keV}\right)^{1.5}}
	+ \frac{4 \times10^4 \left(\frac{0.3{\rm keV}}{h\nu}\right)^{2.5}}{1+8000 \left(T/T_{keV}\right)^2},
	& h\nu > 0.3{\rm keV},
	\end{cases}
	\end{equation*}
	and $\rho=0.001 {\rm g/cm^3}$.
	The heat capacity is given by
	\begin{equation}\label{cvolson}
	\rho C_v = a T_{keV}^3 H \left[1 + \alpha + \left(T+\chi\right) \frac{\partial \alpha}{\partial T}\right],
	\end{equation}
	with
	\begin{equation*}
	\alpha =  \frac{1}{2} e^{-\chi/T} \left(\sqrt{1 + 4e^{\chi/T}} - 1\right), ~~~~~
	\frac{\partial \alpha}{\partial T} = \frac{\chi}{T^2} \left(\alpha - 1/\sqrt{1 + 4e^{\chi/T}}\right).
	\end{equation*}
	In this case, $\chi=0.1 T_{keV}$ and $H=0.1 {\rm cm}$.
	Besides, $a=1$ and $c=1$ are employed in this case.
	The computational domain is $\left[0{\rm cm}, 4.8{\rm cm}\right]$ covered by uniform mesh with cell resolution $\Delta x = 0.0375{\rm cm}$.
	Initially, in the whole domain, the slab is in cold condition with $T=0.01{\rm keV}$.
	All the boundaries are set as reflection boundary conditions.
	There exists a time-independent source term given by $Q\left(x\right)=B\left(0.5{\rm keV}\right) \times e^{-693x^3}$.
	The duration time of the source term is $ct=0 \sim 2$.
	The profiles of material temperature and radiation intensity at different times $ct=2,3,4$ are shown in Figure~\ref{Olson1D},
	and compared with the reference solution obtained by the high order $P_N$ method in \cite{rad-case-olson2020stretched}.
	
	\subsection{Olson 2D case}
\begin{figure}[htbp]
		\centering
		\subfigure{
			\includegraphics[height=6.5cm]{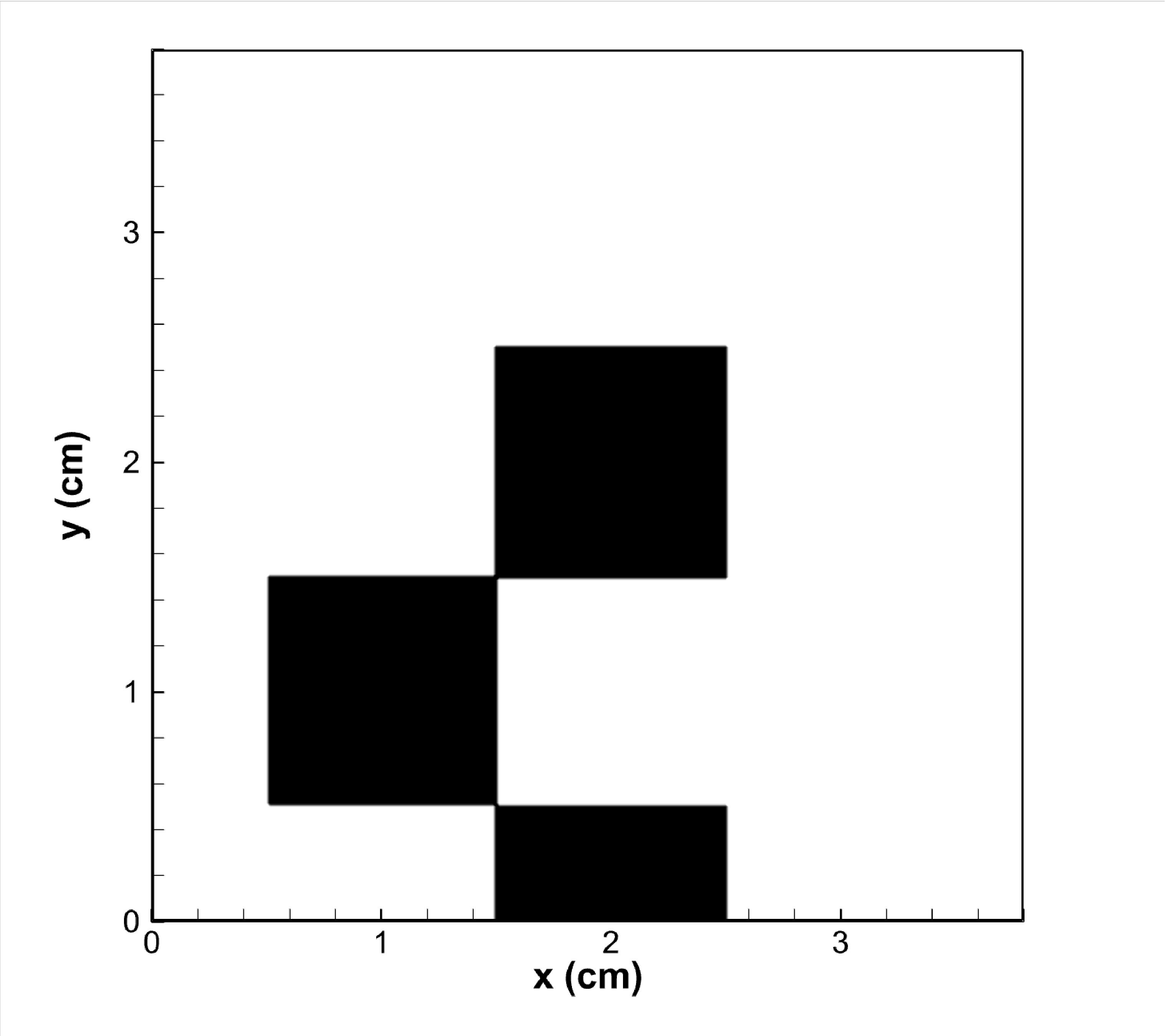}}
		\caption{The distribution of carbon-hydrogen foam (white zone) and aluminum (black blocks) in the domain.}
		\label{Olson2D structure}
	\end{figure}

	In this case, the above Olson case is extended to a 2D problem,
	which is composed of the carbon-hydrogen foam and aluminum blocks in the domain.
	The computational domain is $\left[0{\rm cm},3.8 {\rm cm}\right] \times \left[0{\rm cm},3.8 {\rm cm}\right]$ covered by $285\times285$ uniform mesh.
	The distribution of background materials is shown in Figure~\ref{Olson2D structure},
	where black blocks stand for the optical thicker aluminum material,
	while white zones stand for the optical thinner carbon-hydrogen foam.
	
\begin{figure}[htbp]
		\centering
		\subfigure{
			\includegraphics[height=6.5cm]{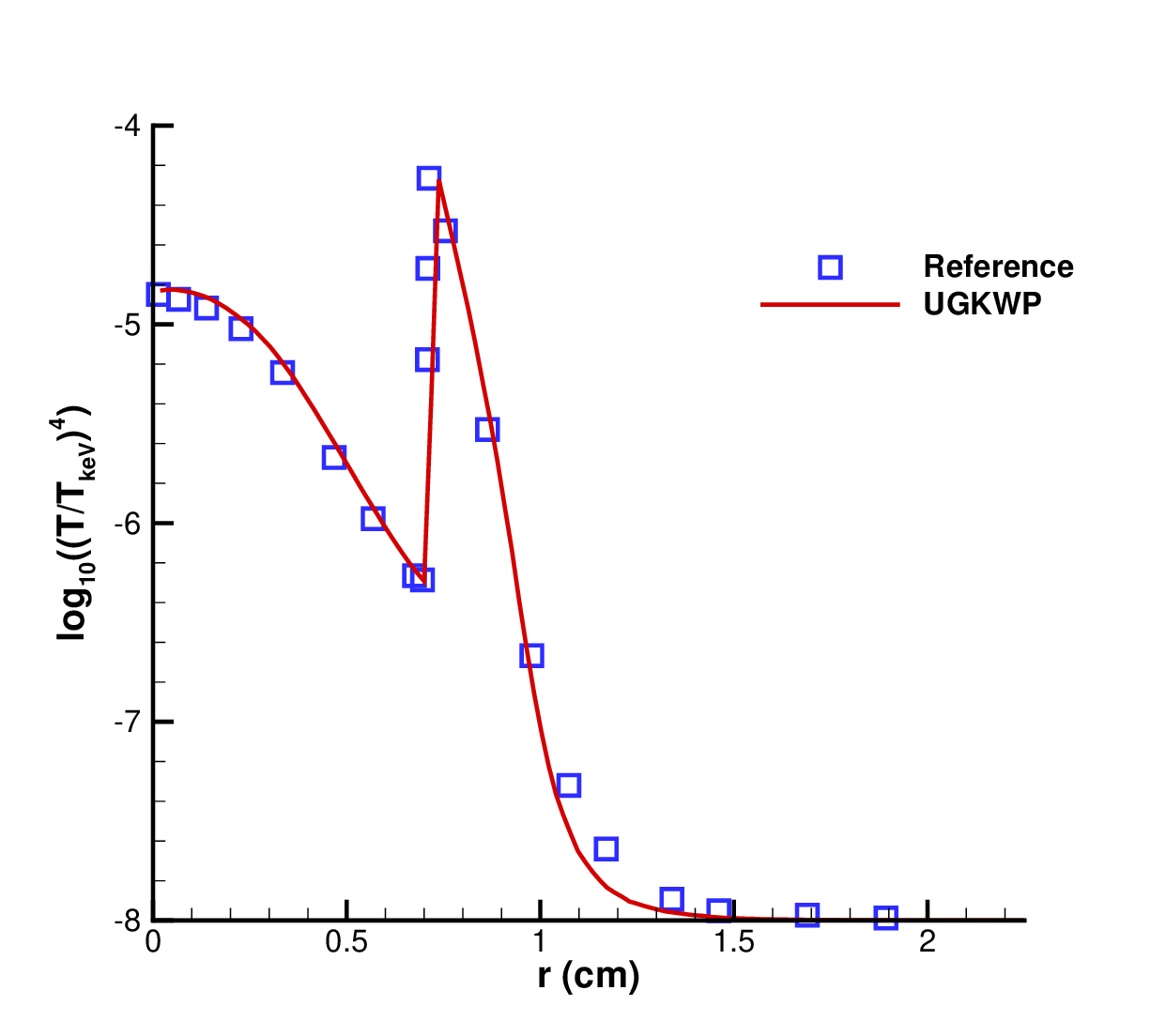}}
		\quad
		\subfigure{
			\includegraphics[height=6.5cm]{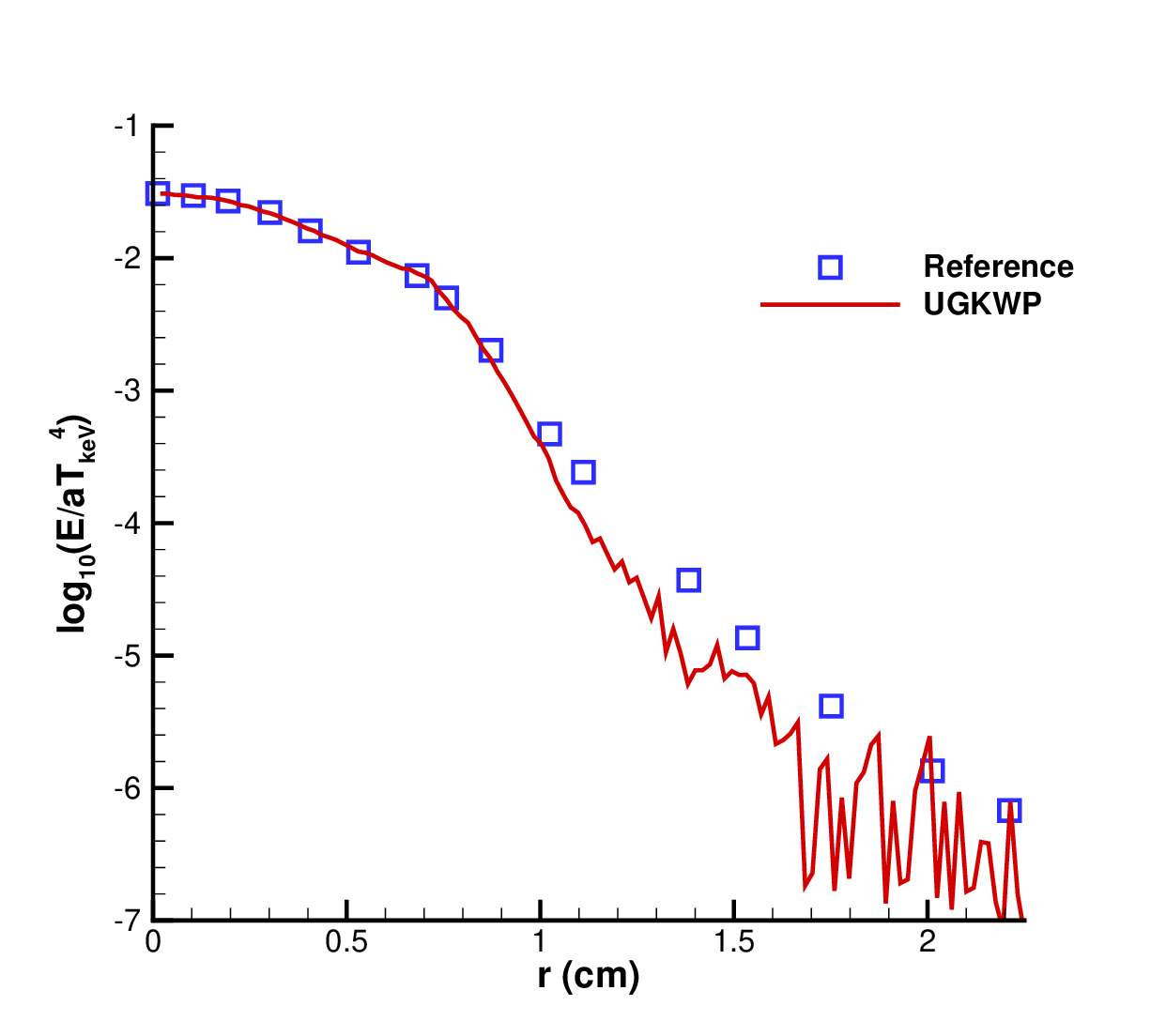}}
		\caption{The distribution of material temperature and radiation intensity along $y=x$ of Olson 2D problem at $ct=3$.}
		\label{Olson2D R}
	\end{figure}

\begin{figure}[htbp]
		\centering
		\subfigure{
			\includegraphics[height=6.5cm]{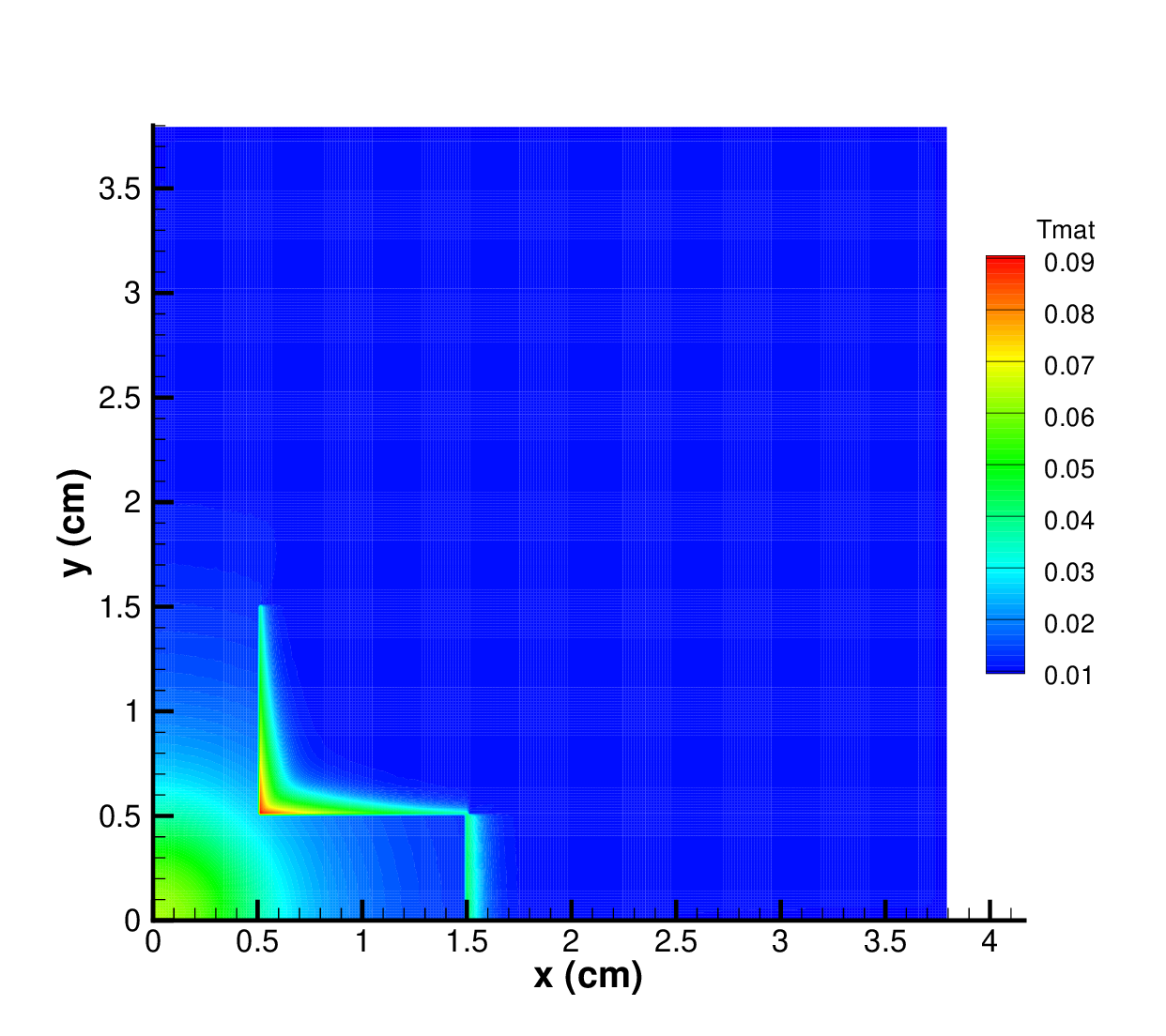}}
		\quad
		\subfigure{
			\includegraphics[height=6.5cm]{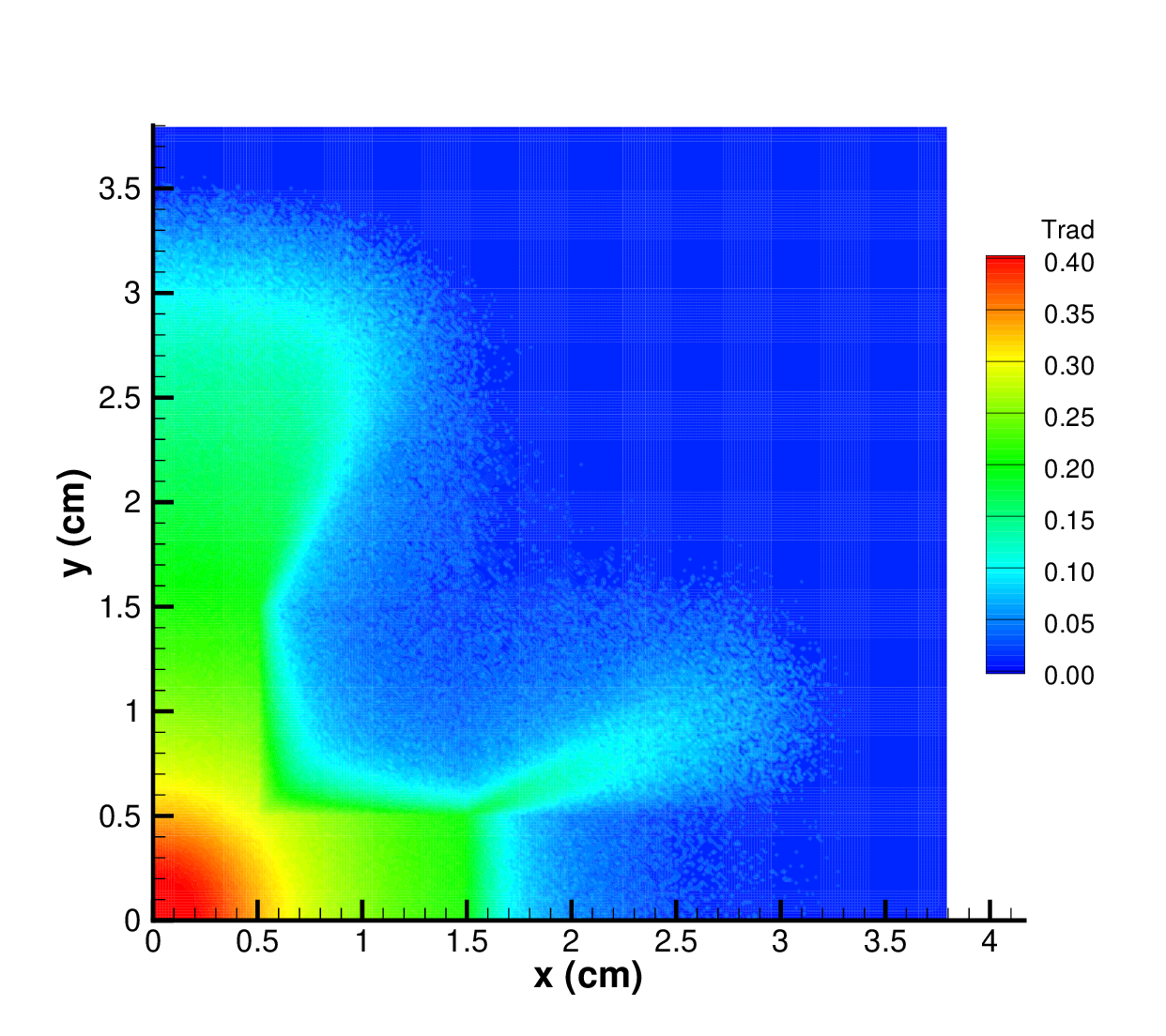}}
		\caption{The snapshots of material temperature and radiation temperature at $ct=3$.}
		\label{Olson2D contour}
	\end{figure}

	The opacity and heat capacity of carbon-hydrogen foam are the same as in the above case. For the aluminum material, the frequency-dependent opacity is calculated by $\sigma = \rho \kappa$ with
	\begin{equation*}
	\kappa=
	\left\{\begin{aligned}
	& \text{Min}\left(10^7, 10^8 \left(T/T_{keV}\right)^2\right),          & h\nu \le 0.01 {\rm keV},   \\
	& \frac{10^7 \left(\frac{0.01{\rm keV}}{h\nu}\right)^2}{1+20 \left(T/T_{keV}\right)^{1.5}},     & 0.01 {\rm keV} < h\nu \le 0.1{\rm keV}, \\
	& \frac{10^7 \left(\frac{0.01{\rm keV}}{h\nu}\right)^2}{1+20 \left(T/T_{keV}\right)^{1.5}} + \frac{10^6 \left(\frac{0.1{\rm keV}}{h\nu}\right)^2}{1+200 \left(T/T_{keV}\right)^{2}},    & 0.1 {\rm {\rm keV}} < h\nu \le 1.5{\rm {\rm keV}},   \\
	& \frac{10^7 \left(\frac{0.01{\rm keV}}{h\nu}\right)^2 \left(\frac{1.5{\rm keV}}{h\nu}\right)^{0.5}}{1+20 \left(T/T_{keV}\right)^{1.5}} + \frac{10^5 \left(\frac{1.5{\rm keV}}{h\nu}\right)^{2.5}}{1+1000 \left(T/T_{keV}\right)^2}, & h\nu > 1.5{\rm keV},
	\end{aligned}\right.
	\end{equation*}
	and $\rho=0.001 {\rm g/cm^3}$, the same with carbon-hydrogen foam. Besides, the heat capacity of aluminum blocks can be evaluated through the Eq.~\eqref{cvolson} with $\chi=0.3 T_{keV}$ and $H=0.5{\rm cm}$. The initial and boundary conditions are the same as 1D case.
	The source term for radiation is added by $Q\left(x\right)=B\left(0.5 {\rm keV} \right) \times e^{-18.7r^3}$ in the simulation, and $r$ stands for the distance from the cell center to the coordinate origin.
	The source term remains in the whole simulation, during $ct=0 \sim 3$.
	The profiles of material temperature and radiation intensity along the diagonal at $ct=3$ are shown in Figure~\ref{Olson2D R}, and compared with the reference solution by $P_N$ method in \cite{rad-case-olson2020stretched}.
	Besides, the contours at $ct=3$ are also given in Figure~\ref{Olson2D contour}.
	
	\section{Conclusion}
	In the paper, the multi-scale UGKWP method is employed to solve the frequency-dependent RTE system.
	Particularly, the frequency space is discretized into several groups to take into consideration the effect of the photon's frequency on the evolution of the whole RTE system.
	The macroscopic system is discretized implicitly and solved by source interaction,
	during which the non-equilibrium photon's transport is evaluated by the explicit tracking of sampled particles in the UGKWP method.
	As a result, each frequency group can find its optimal distribution strategy in the wave-particle decomposition.
	
	In practical problems, the opacity of background material could vary dramatically,
	and therefore the flow regime may have a significant change between two adjacent cells.
	To capture a sharp regime transition from photon's free transport to thermal diffusion,
	it is necessary to construct a multi-scale method.
	In the UGKWP method, two strategies are employed to provide more accurate solutions and avoid numerical oscillation in the region with a steep variation in opacity.
	Firstly, the free streaming time of sampled particle $t_f$ will be reset when this particle passes through the cell interface.
	Secondly, in the zone of thin-thick-joint where the opacity dramatically varies within two adjacent cells,
	a numerical collision time $\tau_{num}$, with the consideration of the material temperature variation between two adjacent cells,
	is employed to increase the portion of stochastic particles in the UGKWP method.
	Several multi-frequency cases covering the optically-thin to the optically-thick regimes have been tested,
	and the results by the UGKWP method agree well with the reference solutions of IMC or $P_N$ method.
	
	\section*{Acknowledgements}
	
	The current research is supported by the National Key R\&D Program of China 2022YFA1004500,
	National Natural Science Foundation of China (Grant No. 12172316),
	and Hong Kong research grant council (16208021, 16301222).
	
	\setcounter{equation}{0}
	\renewcommand\theequation{A.\arabic{equation}}
	
	\section*{Appendix A: Generation of particle frequency from a linear-frequency sampling strategy}
	This appendix will introduce how to determine the frequency value carried by the sampled particles in the UGKWP method under the linear-frequency sampling strategy.
	For the $g$-th group, the cumulative distribution function of $\nu$ can be written as the general form
	\begin{equation}\label{Gformula}
	G\left(\nu\right) = \frac{\int_{\nu_{g-1/2}}^{\nu} B_g\left(\nu\right) \text{d}\nu}{\int_{\nu_{g-1/2}}^{\nu_{g+1/2}} B_g\left(\nu\right) \text{d}\nu}.
	\end{equation}
	With the linear distribution of $B_g\left(\nu\right)$ by Eq.~\eqref{Blinear} and Eq.~\eqref{kg},
	\begin{equation*}
	B_g\left(\nu\right) = B\left(\nu_{g,mid}\right) + k_g \left(\nu - \nu_{g,mid}\right), ~~~\nu_{g,mid} = \frac{\nu_{g-1/2} + \nu_{g+1/2}}{2},
	\end{equation*}
	Eq.~\eqref{Gformula} can be re-written as
	\begin{equation*}
	G\left(\nu\right) = \frac{B\left(\nu_{g,mid}\right) \left(\nu_{g+1/2} - \nu_{g-1/2}\right) + \frac{k}{2} \left(\nu^2 - \nu_{g-1/2}^2\right) - k_g \nu_{g,mid} \left(\nu - \nu_{g-1/2}\right)}{B\left(\nu_{g,mid}\right) \left(\nu_{g+1/2} - \nu_{g-1/2}\right)}.
	\end{equation*}
	From a random number $\eta$, we could obtain the frequency by
	\begin{equation*}
	G\left(\nu\right) = \eta, ~~~ \eta\in U\left(0,1\right),
	\end{equation*}
	i.e., by solving
	\begin{equation}\label{SquareEq}
	a \nu^2 + b \nu + c = 0,
	\end{equation}
	with
	\begin{gather*}
	a = \frac{k_g}{2}, ~~ b = B\left(\nu_{g,mid}\right) - k_g \nu_{g,mid}, \\
	c = k_g\nu_{g-1/2}\nu_{g,mid} - B\left(\nu_{g,mid}\right)\nu_{g-1/2} - \frac{k_g}{2} \nu_{g-1/2}^2 - \eta B\left(\nu_{g,mid}\right) \left(\nu_{g+1/2} - \nu_{g-1/2}\right),
	\end{gather*}
	where the solution to Eq.~\eqref{SquareEq} can be easily obtained.
	
	\setcounter{equation}{0}
	\renewcommand\theequation{B.\arabic{equation}}
	
	\section*{Appendix B: Source iteration method for macroscopic implicit system}
	In this appendix,
	we will introduce the matrix-free source iteration method
	based on the one-dimensional case for solving the RTE system.
	There would be no difficulties in extending the method to two- and three-dimensional cases.
	Specifically, the angular space is reduced to $\mu \in [-1,1]$ in one dimensional case,
	and Eq.~\eqref{macroSysFinal} for $i$-th cell can be written as
	\begin{equation}\label{macroSysAppendix}
	\left\{
	\begin{array}{c}
	\rho_{g,i}^{n+1} = \rho_{g,i}^n
	+ \frac{\Delta t}{\Delta x} \left(F^{eq}_{g,i-\frac{1}{2}} - F^{eq}_{g,i+\frac{1}{2}}\right)
	+ \frac{\Delta t}{\Delta x} \left(F^{fr,wave}_{g,i-\frac{1}{2}} - F^{fr,wave}_{g,i+\frac{1}{2}}\right)                                \\
	~~~+ \frac{w_{g,i}^{fr,p}}{V_{i}}
	+ \frac{c \Delta t}{\epsilon^2} \left(2\sigma_{g,i}^{n+1} B_{g,i}^{n+1} - \sigma_{g,i}^{n+1} \rho_{g,i}^{n+1} \right), ~~~ g=1,...,G, \\
	C_v T_i^{n+1} = C_v T_i^{n} + \sum_{g=1}^{G} \frac{\Delta t}{\epsilon^2} \left(\sigma_{g,i}^{n+1} \rho_{g,i}^{n+1} - 2\sigma_{g,i}^{n+1} B_{g,i}^{n+1} \right).
	\end{array}
	\right.
	\end{equation}
	The equilibrium flux $F_{g}^{eq}$ can be calculated by Eq.~\eqref{FeqFormula}, which indicates
	\begin{align*}
	F^{eq}_{g,i-\frac{1}{2}}
	& = C_{2,i-\frac{1}{2}} \int_{-1}^{1} \frac{c}{\epsilon} \mu
	\left( \frac{1}{2} \frac{c}{\epsilon} \mu B_{g,x,i-\frac{1}{2}} \right) \text{d}\mu                                             \\
	& = C_{2,i-\frac{1}{2}} \frac{c^2}{\epsilon^2} \frac{1}{2}\frac{B_{g,i} - B_{g,i-1}}{\Delta x} \int_{-1}^{1} \mu^2 \text{d}\mu \\
	& = C_{2,i-\frac{1}{2}} \frac{c^2}{\epsilon^2} \frac{1}{3}\frac{B_{g,i} - B_{g,i-1}}{\Delta x},
	\end{align*}
	and similarly,
	\begin{align*}
	F^{eq}_{g,i+\frac{1}{2}}
	& = C_{2,i+\frac{1}{2}} \frac{c^2}{\epsilon^2} \frac{1}{3}\frac{B_{g,i+1} - B_{g,i}}{\Delta x},
	\end{align*}
	where $C_{2,i-\frac{1}{2}}, C_{2,i+\frac{1}{2}}$ are the coefficients calculated by Eq.~\eqref{C2Formula}.
	As a result, the equilibrium flux terms in Eq.~\eqref{macroSysAppendix} can be written as,
	\begin{equation*}
	\frac{\Delta t}{\Delta x} \left(F^{eq}_{g,i-\frac{1}{2}} - F^{eq}_{g,i+\frac{1}{2}}\right)
	= \frac{\Delta t}{\Delta x} \left[ C_{2,i-\frac{1}{2}} \frac{c^2}{\epsilon^2} \frac{1}{3}\frac{B_{g,i} - B_{g,i-1}}{\Delta x} - C_{2,i+\frac{1}{2}} \frac{c^2}{\epsilon^2} \frac{1}{3}\frac{B_{g,i+1} - B_{g,i}}{\Delta x} \right]
	\overset{def}{=} K_g B_{g,i} - M_g,
	\end{equation*}
	with
	\begin{equation*}
	K_g = \frac{c^2}{\epsilon^2} \frac{1}{3} \frac{\Delta t}{\Delta x^2}
	\left(C_{2,i-\frac{1}{2}} + C_{2,i+\frac{1}{2}}\right), ~~~
	M_g = \frac{c^2}{\epsilon^2} \frac{1}{3} \frac{\Delta t}{\Delta x^2}
	\left(C_{2,i-\frac{1}{2}} B_{g,i-1} + C_{2,i+\frac{1}{2}} B_{g,i+1}\right).
	\end{equation*}
	It should be noted that the opacity at the interface $\sigma_{i-\frac{1}{2}}$ is evaluated by
	\begin{equation*}
	\sigma_{i-\frac{1}{2}} = \frac{2\sigma^{l}_{i-\frac{1}{2}}\sigma^{r}_{i-\frac{1}{2}}}{\sigma^{l}_{i-\frac{1}{2}} + \sigma^{r}_{i-\frac{1}{2}}},
	\end{equation*}
	where $\sigma^{l}_{i-\frac{1}{2}}$ and $\sigma^{r}_{i-\frac{1}{2}}$ are the values on the left and right sides of the interface, respectively,
	which are obtained by spatial reconstruction of $\sigma$.
	As a result, Eq.~\eqref{macroSysAppendix} can be re-written as
	\begin{equation}\label{macroSysAppendixWithKM}
	\left\{
	\begin{array}{c}
	\rho_{g,i}^{n+1} = \rho_{g,i}^n
	+ \left(K_g B_{g,i} - M_g \right)
	+ \frac{\Delta t}{\Delta x} \left(F^{fr,wave}_{g,i-\frac{1}{2}} - F^{fr,wave}_{g,i+\frac{1}{2}}\right)                                \\
	~~~+ \frac{w_{g,i}^{fr,p}}{V_{i}}
	+ \frac{c \Delta t}{\epsilon^2} \left(2\sigma_{g,i}^{n+1} B_{g,i}^{n+1} - \sigma_{g,i}^{n+1} \rho_{g,i}^{n+1} \right), ~~~ g=1,...,G, \\
	C_v T_i^{n+1} = C_v T_i^{n} + \sum_{g=1}^{G} \frac{\Delta t}{\epsilon^2} \left(\sigma_{g,i}^{n+1} \rho_{g,i}^{n+1} - 2\sigma_{g,i}^{n+1} B_{g,i}^{n+1} \right),
	\end{array}
	\right.
	\end{equation}
	which can be solved by the source iteration method.
	Particularly, in the $k$-th iteration step, the following equivalent system is adopted
	\begin{equation}\label{macroSysAppendixIte}
	\left\{
	\begin{array}{c}
	\rho_{g,i}^{k+1} = \rho_{g,i}^n
	+ \left(K_g^{k} B_{g,i}^{k} - M_g^{n+1,k} \right)
	+ \frac{\Delta t}{\Delta x} \left(F^{fr,wave,k}_{g,i-\frac{1}{2}} - F^{fr,wave,k}_{g,i+\frac{1}{2}}\right)                      \\
	~~~+ \frac{w_{g,i}^{fr,p}}{V_{i}}
	+ \frac{c \Delta t}{\epsilon^2} \left(2\sigma_{g,i}^{k} B_{g,i}^{k} - \sigma_{g,i}^{k} \rho_{g,i}^{k+1} \right), ~~~ g=1,...,G, \\
	C_v T_i^{k+1} = C_v T_i^{n} + \sum_{g=1}^{G} \frac{\Delta t}{\epsilon^2} \left(\sigma_{g,i}^{k} \rho_{g,i}^{k+1} - 2\sigma_{g,i}^{k} B_{g,i}^{k} \right),
	\end{array}
	\right.
	\end{equation}
	and the procedures are summarized in the following.
	
	\begin{algorithm}[htb]
		\caption{Source iteration for solving Eq.~\eqref{macroSysAppendixWithKM}: find $\rho_{g,i}^{n+1}$, $g=1,...,G$, $T_i^{n+1}$.}
		\begin{algorithmic}[1]
			\STATE Start		
			\STATE Set the initial iteration value: $k=0$, $\rho_{g,i}^{0}=\rho_{g,i}^{n}$, $B_{g,i}^{0}$ based on $T_i^{n}$.
			\WHILE{ $|\rho_{g,i}^{k+1} - \rho_{g,i}^{k}|_{\text{max}} > \epsilon_{\rho} $,
				or, $|T_i^{k+1} - T_i^{k}|_{\text{max}} > \epsilon_{T} $}
			\STATE Compute the coefficients, i.e., $\sigma_{g,i}^{k}$, $K_{g,i}^{k}$, $M_{g,i}^{k}$, based on $\rho_{g,i}^{k}$, $B_{g,i}^{k}$.
			\STATE Calculate the flux $F_{g,i\pm\frac{1}{2}}^{fr,wave,k}$, based on $\rho_{g,i}^{k}$.
			\STATE Update $\rho_{g,i}^{k+1}$ based on $M_{g,i}^{k}$, $F_{g,i\pm\frac{1}{2}}^{fr,wave,k}$, $w_{g,i}^{fr,p}$, etc.
			\STATE Update $T_i^{k+1}$ and associated $B_{g,i}^{k+1}$ based on $\rho_{g,i}^{k+1}$, $B_{g,i}^{k}$, etc.
			\STATE $k=k+1$.
			\ENDWHILE		
			\STATE End
		\end{algorithmic}
	\end{algorithm}

	\normalem
	\bibliographystyle{unsrt}%
	\bibliography{reference}
	
\end{document}